\documentclass{JHEP3}

\usepackage{epsfig,multicol,bbm}
\usepackage{graphicx}
\usepackage{dcolumn}
\usepackage{amsmath}
\usepackage{enumerate}

\newcommand{\mone}{$^{-1}$}

\newcommand{\cub}{$^3$}

\newcommand{\lcdm}{$\Lambda$CDM}
\newcommand{\lcdmn}{$\nu$\lcdm}
\newcommand{\omegam}{$\Omega_{\rm m}$}

\newcommand{\smnu}{$\Sigma \, m_\nu$}
\newcommand{\smnut}{$\Sigma \, m_\nu=(0,0.17,0.3,0.53)$ eV}

\newcommand{\CAMB}{\textsc{CAMB}}
\newcommand{\CLASS}{\textsc{CLASS}}
\newcommand{\Halofit}{\textsc{Halofit}}
\newcommand{\LensPix}{\textsc{LensPix}}
\newcommand{\gadgetthree}{\textsc{gadget-3}}
\newcommand{\subfind}{\textsc{subfind}}
\newcommand{\sun}{\odot}
\newcommand{\msun}{$M_{\sun}$}

\newcommand{\hmone}{$\,h^{-1}$}

\newcommand{\planck}{{\it Planck}}

\newcommand{\eg}{\emph {e.g.} }
\newcommand{\ie}{\emph {i.e.} }

\title{DEMNUni: ISW, Rees-Sciama, and weak-lensing in the presence of  massive neutrinos}

\author{Carmelita Carbone\\ 
INAF--Osservatorio Astronomico di Brera, Via Bianchi 46, 23807, Merate (MI), Italy\\
INFN--Sezione di Bologna, viale Berti Pichat 6/2, I-40127, Bologna (BO), Italy\\
E-mail: \email{carmelita.carbone@brera.inaf.it}}

\author{Margarita Petkova\\
Excellence Cluster Universe, Boltzmannstr. 2, D-85748 Garching,
Germany\\
Universit\"ats-Sternwarte, Fakult\"at f\"ur Physik,
Ludwig-Maximilians Universit\"at M\"unchen, Scheinerstr. 1, D-81679
M\"unchen, Germany\\
E-mail: \email{mpetkova@usm.lmu.de}}

\author{Klaus Dolag\\
Universit\"ats-Sternwarte, Fakult\"at f\"ur Physik,
Ludwig-Maximilians Universit\"at M\"unchen, Scheinerstr. 1, D-81679
M\"unchen, Germany\\
Max-Planck-Institut f\"ur Astrophysik, Karl-Schwarzschild
Strasse 1, Garching bei M\"unchen, Germany\\
E-mail: \email{kdolag@mpa-garching.mpg.de}}

\abstract{We present, for the first time in the literature, a full reconstruction of
the total (linear and non-linear) ISW/Rees-Sciama effect in the presence of
massive neutrinos, together with its cross-correlations with
CMB-lensing and weak-lensing signals. The present analyses make use of
all-sky maps extracted via ray-tracing 
across the gravitational potential distribution provided by the ``Dark
Energy and Massive Neutrino Universe'' (DEMNUni) project, a set of large-volume,
high-resolution cosmological N-body simulations, where neutrinos are
treated as separate collisionless particles. We correctly recover, at
$1-2$\% accuracy, the linear predictions from \CAMB. Concerning the
CMB-lensing and weak-lensing signals, we also recover, with similar
accuracy, the signal predicted by Boltzmann codes, once non-linear
neutrino corrections to \Halofit\ are accounted for. Interestingly, in
the ISW/Rees-Sciama signal, and its cross correlation with lensing, we
find an excess of power with respect to the massless case, due to free
streaming neutrinos, roughly at the transition scale between the
linear and non-linear regimes. The excess is $\sim 5-10$\% at
$l\sim 100$ for the ISW/Rees-Sciama auto power spectrum, depending on
the total neutrino mass $M_\nu$, and becomes a
factor of $\sim 4$ for $M_\nu=0.3$ eV, at $l\sim 600$, for the
ISW/Rees-Sciama cross power with CMB-lensing. This effect should be
taken into account for the correct estimation of the CMB temperature
bispectrum in the presence of massive neutrinos.}


\begin{document}
\section{Introduction}
\label{intro}
The Standard Model of particle physics predicts the existence of three active 
massless neutrino species: the electron ($\nu_{\rm e}$), muon ($\nu_\mu$) and tau ($\nu_\tau$) 
neutrinos. However, the discovery of
lepton flavour oscillations has suggested that neutrinos are massive
particles, fixing the lower limit of the sum of neutrino masses to
\smnu$\equiv m_{\nu_e}+m_{\nu_\mu}+m_{\nu_\tau}\gtrsim 0.06$ eV\footnote{
More specifically, \smnu\ must be greater than approximately $0.06$ eV in the 
normal hierarchy scenario and $0.1$ eV in the degenerate hierarchy.
}~\cite{lesgourgues06,lesgourgues12,lesgourgues14,
lesgourgues13}. This implies that, after becoming
non-relativistic, neutrino free-stream with large thermal
velocities that suppress the growth of neutrino densities 
perturbations on scales smaller than the so-called ``free-streaming
length'', $\lambda_{\rm fs}(z,m_\nu) \simeq 8.1\,H_0\,(1+z)/H(z)\,(1\,\rm eV/m_\nu)\, \rm Mpc/h$,
where $m_\nu$ is the mass of the single neutrino species, $H(z)$ is
the so-called Hubble parameter, and $H_0$ the Hubble constant,
$H(z=0)$. As a consequence, due to gravitational backreaction effects,
also the evolution of cold dark matter (CDM) and
baryon~\cite{rossi14} densities is altered, and the  
total matter power spectrum is largely suppressed at scales $\lambda<<\lambda_{\rm fs}$ .

Moreover, given the present mass constraints, neutrinos become non-relativistic after the epoch of 
recombination, and, accordingly, modify the radiation density contribution. 
The transition from the relativistic to the non-relativistic regimes postpones the matter radiation equality for a given value of
\omegam$\,h^2$  
(where \omegam\ is the ratio, at $z=0$, between the matter density of the Universe and the critical 
density, $\rho_{\rm c}$, and $h$ the Hubble constant $H_0$ in units of 100 km 
s\mone Mpc\mone), and modifies the background evolution, slightly affecting 
the properties of the primary cosmic microwave background (CMB)
anisotropies. 

In addition, along their travel from the last scattering surface to the observer,
CMB photons undergo also secondary anisotropies, in particular they are red/blue-shifted as
they cross growing/decaying gravitational potential wells. 
This effect is called the late Integrated Sachs-Wolfe effect (ISW), as it was first described by
Sachs and Wolfe in 1967~\cite{Sachs67}. During the matter
dominated era, the two effects of background expansion and gravitational attraction
compensate each other so that the total linear gravitational
potential, $\Phi$, produced by the Large Scale Structure (LSS) distribution in the
Universe, is constant in time, and the ISW effect, which depends on the
time derivative $\dot{\Phi}$, vanishes. In contrast, during \eg the Dark Energy (DE) dominated
era, the background expansion rate of the Universe increases and these
two effects do not compensate anymore, causing the
decaying of the gravitational potential perturbations. In this case,
$\dot{\Phi}$ is no longer vanishing: a CMB photon passing through an overdense
region gains more energy falling into the potential
well with respect to the energy lost while climbing out of it; a
CMB photon passing through an underdense region loses more energy climbing the potential hill than
the energy gained during its descent. Therefore, overdense
regions correspond to hotter spots in the CMB sky map, and underdense regions to colder ones.

In the linear regime, the total effect is represented by an increase
of the photon temperature power spectrum on very large scales, which
has actually been detected via full-sky CMB probes,
\eg\ Planck~\cite{Planck13,Planck15}, or via the cross-correlation of the CMB temperature with LSS data~\cite{2012MNRAS.426.2581G,2016arXiv160403939S,Granett2009}. 

However, besides dark energy, also the cosmological background of
massive neutrinos produces a non-vanishing ISW effect. In fact, the neutrino free-streaming
makes the gravitational potential to evolve in time, producing a net
$\dot{\Phi} \neq 0$ even in the absence of a recent accelerated background expansion.
As shown in~\cite{Lesgourgues_etal_2008}, neutrino velocities generate an
excess of ISW effect at high redshifts $z$, due to their
impact on the linear growth factor. Unfortunately, this effect is not
directly observable, since its detection would require very precise data at large $l$ and high
redshifts, where the late ISW effect is masked by primordial
temperature anisotropies. 

The non-linear growth of density perturbations modifies the previous picture,
producing additional temperature perturbations which give rise to the
so-called ``Rees-Sciama'' (RS) effect, directly related to the
momentum density in the non-linear regime ({\emph e.g.}
\cite{RS68,Cooray2002,Seljak96,Spergel98,Merkel_isw}, and references therein).
In fact, the accelerated non-linear growth of structure
increases the depth of the potential wells in overdense
regions, resulting in a reduction of the total CMB temperature, with
respect to the linear case. This partially cancels the ISW effect in
hotter regions. On the contrary, the RS effect increases the ISW
effect in underdense regions, since the saturation of the density contrast in
voids further suppresses the growth of the gravitational
perturbations. Also in this case, massive neutrinos have an impact,
since they alter the RS effect in a scale- and redshift-dependent way,
owing to the neutrino free-streaming scale, $\lambda_{\rm fs}$.

Besides the ISW and RS effects, CMB photons undergo also the gravitational
lensing~\cite{Lewis06} and the Sunyaev-Zeldovich (SZ)
effects~\cite{SZ:1969} generated by LSS. Massive neutrinos alter the
related auto-~\cite{Roncarelli_etal_2015} and
cross-correlation functions, with different impacts at
the linear and non-linear levels, mainly depending on their effect on
the total matter power spectrum and cluster number counts. 

In this work, we present for the first time in the literature a full reconstruction of
the total (linear and non-linear) ISW-RS effect in the presence of
massive neutrinos, together with its cross-correlations with
CMB-lensing and weak-lensing signals. 
Previous works \cite{Cai_isw,Jubilee_isw} have provided a similar
reconstruction in the standard $\Lambda$CDM massless case. 

The present analyses make use of all-sky maps extracted via ray-tracing
across the ``Dark Energy and Massive Neutrino Universe'' (DEMNUni)
simulations, which are the largest N-body simulations to date with a
particle neutrino component. At present, as we explain in more details
in \S~\ref{demnuni} below, these simulations are characterised by a
baseline $\Lambda$CDM cosmology to which we add neutrinos with
different total masses, $M_\nu\equiv$\smnu. In the next future, we plan to extend
the DEMNUni set with the inclusion of an evolving dark energy
background, with different equations of state $w$, in order to study
the degeneracy between $M_\nu$ and $w$ at the non
linear level. 

This paper is organised as follows. In \S~\ref{demnuni} we
present the DEMNUni simulations. In
\S~\ref{map-making}, we explain the map-making procedure, in
\S~\ref{auto-power}  we present
the ISW-RS and lensing signals extracted 
from the simulations, focusing the discussion on the angular auto
power spectra, and presenting the cross-correlation signals in \S~\ref{cross-power}. Finally,
in \S~\ref{conclu} we draw our main conclusions. 

\section{The DEMNUni simulations}
\label{demnuni}
The DEMNUni simulations have been conceived for the analysis of different
probes, like galaxy surveys and CMB data,
and their cross-correlations, in the presence of massive neutrinos. In particular,
in order to investigate simultaneously the neutrino impact on
different CMB secondary anisotropies, \eg the ISW-RS and weak-lensing
effects, we
have produced a set of simulations with a volume big enough
to include the very large scale perturbation modes, and, at the same time, with a good mass resolution to
investigate the effects of small-scale non-linearities and neutrino
free-streaming.  Moreover, for the accurate reconstruction of the
light-cone back to the starting redshift of the simulations, we have
assumed an output time-spacing small enough that possible systematic
errors, due to the interpolation between neighbouring redshifts along
the line of sight, result to be negligible.

The DEMNUni simulations have been performed using the tree particle 
mesh-smoothed particle hydrodynamics (TreePM-SPH) code
\gadgetthree~\cite{springel05}, specifically modified by~\cite{viel10} to   
account for the presence of massive neutrinos. This modified version of
\gadgetthree\ follows the evolution of CDM and neutrino particles,
treating them as two separated collisionless species. Given the
relatively high velocity dispersion, neutrinos have a characteristic clustering scale larger than 
the CDM one, allowing to save computational time by neglecting the calculation of the 
short-range gravitational force. This results in a different spatial resolution for the 
two components, which for neutrinos is fixed by the PM grid (that we have
chosen to be eight times larger than the particle number), while for
CDM particles is about one order of magnitude higher.

The DEMNUni set of simulations has a starting redshift $z_{in}=99$,
and is characterised by a comoving volume of
(2 \hmone\ Gpc)\cub\, filled with 2048\cub\ dark matter
particles and, where present, 2048\cub\ neutrino particles. Given the large
amount of memory required by the simulations, baryon physics is not included. 
The authors in Ref.~\cite{vanDaalen_etal_2011} found baryon effects to
be independent of cosmological parameters, suggesting that they 
are also independent of the neutrino mass; therefore, our choice
should not affect the results presented in this work. This is
supported also by~\cite{Bird_etal2011}, where the authors show that
the neutrino induced suppression in the matter power spectrum is very much the
same also when neutrinos are considered in the presence of
baryons. Moreover, since we are seeking deviations (due to massive
neutrinos) from a fiducial reference $\Lambda$CDM model in terms of
$P(k)$ ratios,  we expect that baryon feedback will cancel out in
this case, and that additional effects produced  by the interplay of
neutrinos with baryon physics should be higher order in both (for
further details see~\cite{Castorina_etal_2015}). 

We have produced a total of four different simulations, choosing the
cosmological parameters according to the
\planck\ results~\cite{planck14cp}, namely a flat \lcdm\ model
generalised to a \lcdmn\  
framework by changing only the value of the sum of the three active neutrino masses $M_\nu=$\smnut, 
respectively, and keeping fixed \omegam\ and the amplitude of primordial curvature 
perturbations $A_{\rm s}$. 

The simulations are characterised by a softening length
$\varepsilon=20$\hmone\ Kpc, and have been run on the Fermi IBM BG/Q
supercomputer at CINECA\footnote{http://www.cineca.it/},
Italy, employing about 1 Million cpu-hrs per
simulation (including the production of halo and sub-halo catalogues). 
For each simulation we have produced 62 output logarithmically equispaced in the scale factor 
$a=1/(1+z)$, in the redshift interval $z=0-99$, 49 of which lay
between $z=0$ and $z=10$. For each of the 62 output times, we have dumped
on-the-fly a particle snapshot composed by both CDM and neutrino
particles, a three-dimensional (3D) grid of the
gravitational potential, $\Phi$, with side size $L_{\rm
  box}=2$\hmone\ Gpc and a mesh of 4096\cub\ cells, and a 3D grid of the time
derivative $\dot{\Phi}$, with same dimensions and resolution, for a
total of about 90 TB of data per simulation. Finally, in order to build halo catalogues,  
we have post-processed each of the 62 particle snapshots with the
friends-of-friends (FoF) algorithm, included in
\gadgetthree~\cite{springel01,dolag09}, 
setting to 32 the minimum number of particles, thus fixing the halo
minimum mass to $M_{\rm FoF}\simeq2.5\times
10^{12}$\hmone\msun. Finally, the FoF catalogues have been processed via the
\subfind\ algorithm (also included in \gadgetthree\ ) so that the
initial FoF parent halos are split into multiple sub-halos, with the
result of an increase in the total number of identified objects and of
a lower minimum mass limit (for further details see~\cite{Castorina_etal_2015}). 

\begin{figure*}[!ht]
\begin{center}
\setlength{\tabcolsep}{0.01pt}
\begin{tabular}{c c}
\includegraphics[width=0.5\textwidth]{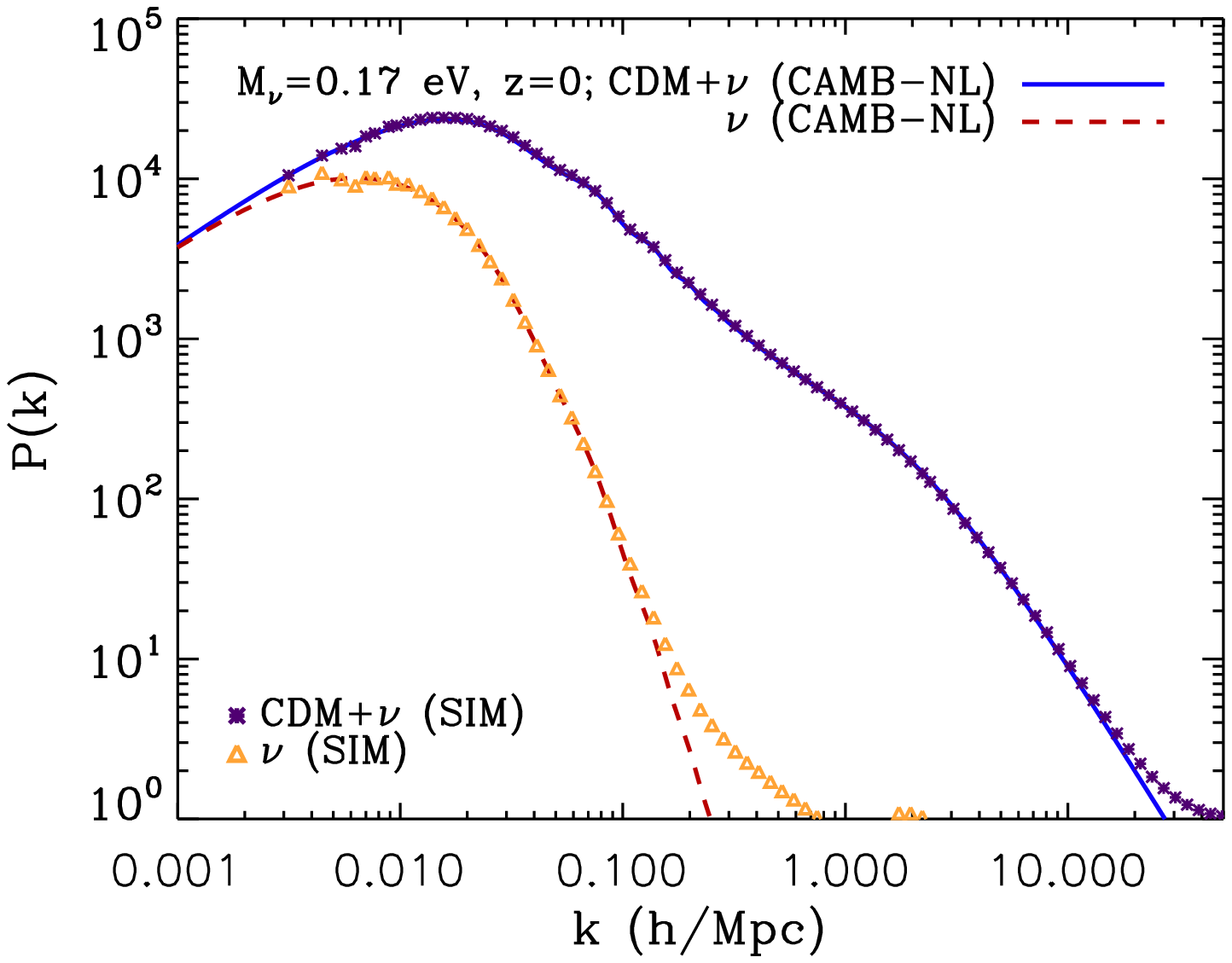}&
\includegraphics[width=0.5\textwidth]{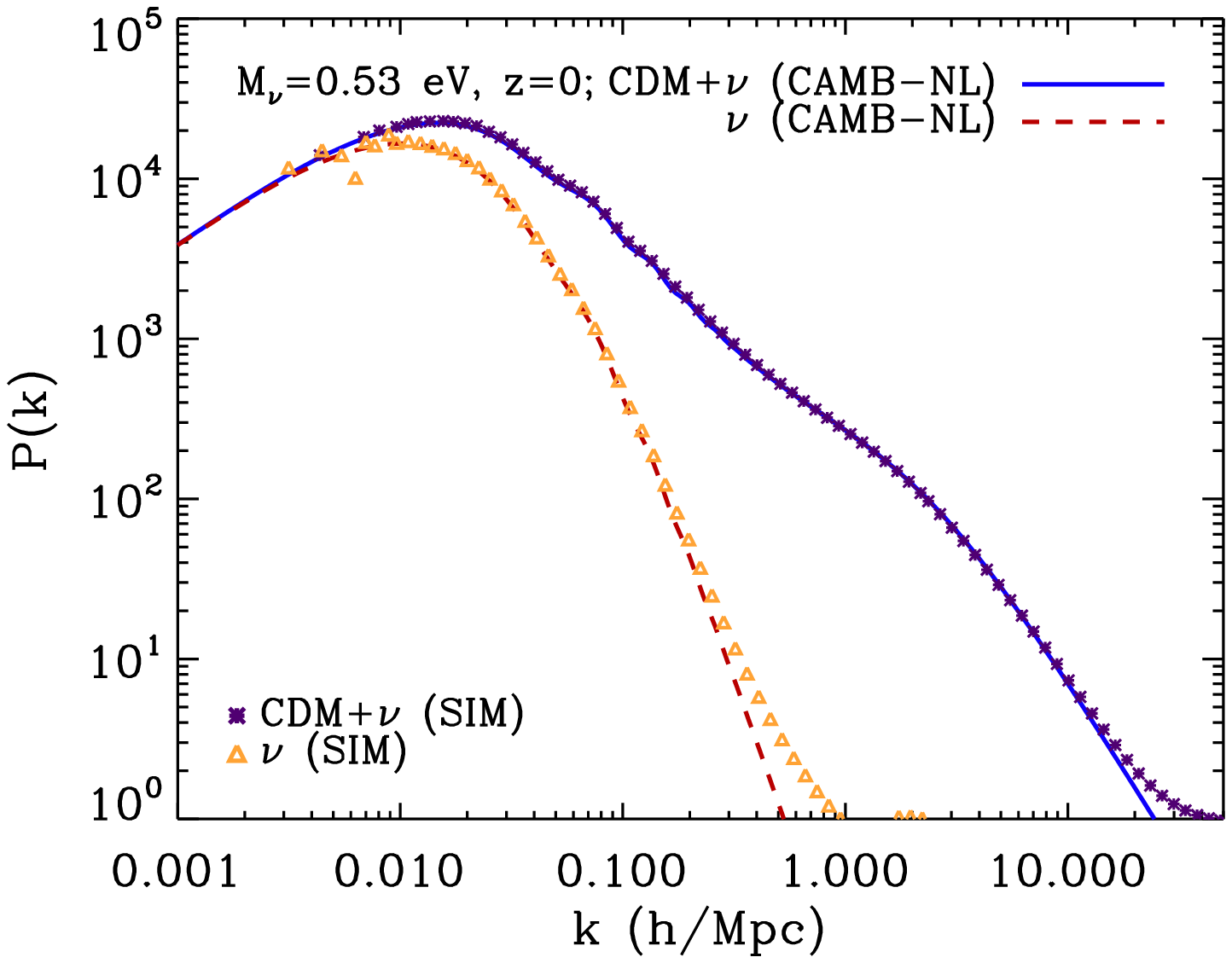}\\
\includegraphics[width=0.5\textwidth]{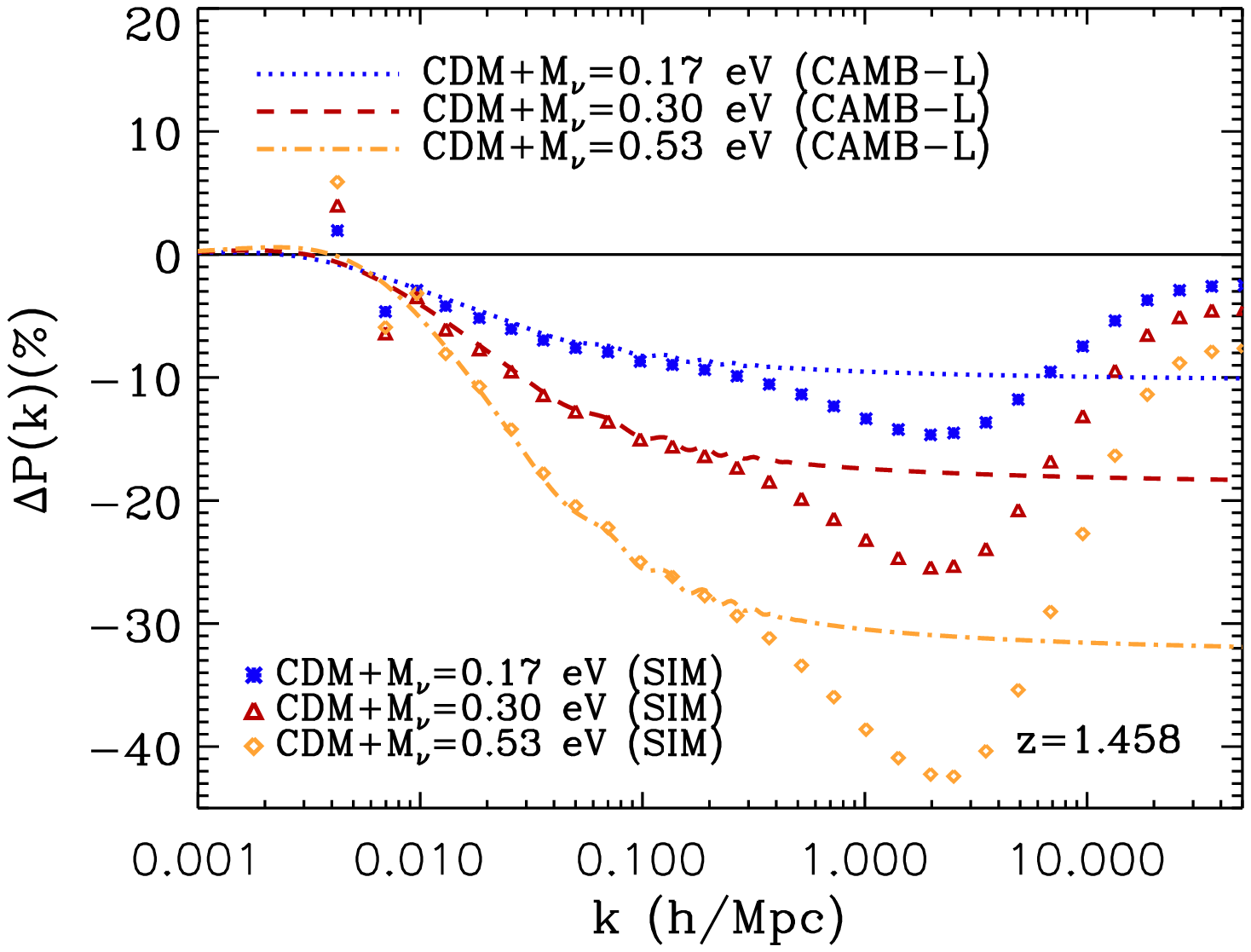}&
\includegraphics[width=0.5\textwidth]{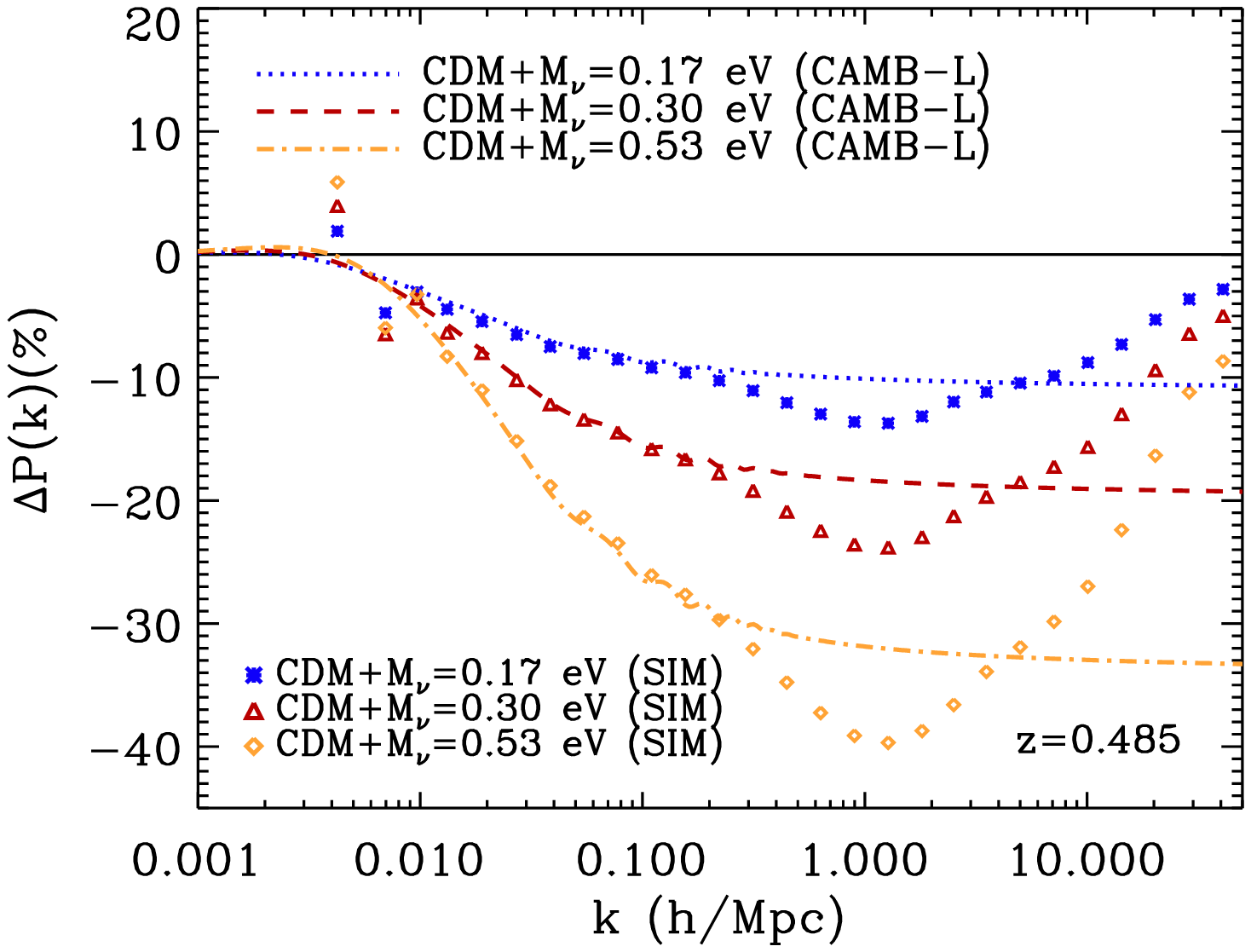}
 \end{tabular}
\end{center} 
\caption{Top panel: the total matter $P(k, z=0)$ for $M_\nu=0.17,
  0.53$ eV, measured from the simulations (violet stars) compared to
  the non-linear (NL) total matter $P(k)$ extracted  from
  \CAMB\ (solid blue line) combined with the \Halofit\ \cite{Smith_etal_2003} non-linear
  corrections, including the neutrino contribution from \cite{Bird_etal2011}. Orange
  triangles and red dashed lines represent the corresponding
  neutrino $P(k)$ from simulations and \CAMB, respectively.
Bottom panel: ratios of the simulated total matter $P(k)$ for
$M_\nu=0.17, 0.30, 0.53$ eV (stars, triangles, diamonds,
respectively) wrt the simulated $P(k, M_\nu=0 \;{\rm eV})$ at $z=0.485,
1.458$. The dotted blue, dashed red, and dot-dashed orange lines
represent the linear (L) expectations from \CAMB.}
\label{pk}
\end{figure*}

In Fig.~\ref{pk} the power spectra extracted from the simulations 
for two total neutrino masses are shown (upper panel), together with their ratios
with respect to the $\Lambda$CDM case (lower panel), for different neutrino
masses and at two different redshifts. Worth of note is the range of scales sampled by the
simulations, more than three order of magnitudes, thanks both to the
peculiar large volume and mass resolution characterising the DEMNUni
set. The lower panel of Fig.~\ref{pk} shows the well-known non-linear
damping caused by massive neutrinos on the total matter power
spectrum. Our findings recover previous results in the literature
~\cite{2014JCAP...03..011V, 2014JCAP...02..049C, 2009JCAP...05..002B,
  2008JCAP...08..020B, 2010JCAP...01..021B, 2013MNRAS.428.3375A,
  2013JCAP...03..019V, Massara2014,
Wagner2012, Upadhye2014}, in particular the excess of power
suppression with respect to the linear theoretical expectations. As
we will show in the next Sections, the non-linear behaviour of the
total matter power spectrum proportionally affects the CMB-lensing
and weak-lensing potentials from LSS, and the larger the total neutrino
mass $M_\nu$ is, the greater its impact on lensing quantities is. On
the contrary, we will show that, for the ISW-RS effect, lighter
neutrinos produce, at intermediate scales, a larger effect on
$\dot{\Phi}$, due to their higher thermal velocities, $v_{\rm
  th}(z,m_\nu) \simeq 158\,(1+z)\,(1 {\rm eV}/m_\nu)\,{\rm km/s}$, and smaller
free-streaming comoving wave number in Fourier space, $k_{fs} \equiv 2\pi/\lambda_{fs}/(1+z)$.

\section{ISW-RS and weak-lensing potential reconstruction}
\label{map-making}
The temperature anisotropies induced by the total (linear and
non-linear) ISW-RS effect, in a direction ${\bf \hat{n}}$ on the sky, can be
computed as the integral of the time derivative of the physical
peculiar gravitational
potential, $\dot{\Phi}$, along the line of sight from the last
scattering (LS) surface to the present epoch at $t_0$~\cite{Sachs67}.
\begin{equation}
\label{eq1}
\Delta T ({\bf \hat{n}}) = \frac{2}{c^2}\bar{T_0}\int_{t_{\rm LS}}^{t_0}
\dot{\Phi}(t,{\bf \hat{n}}) \, dt,
\end{equation}
where $t$ is the cosmic time, $t_{\rm LS}$ the age of the Universe at
the LS surface, $\bar{T_0}=2.7255 {\rm K}$ the today CMB temperature,
and $c$  the speed of light. Equation~(\ref{eq1}) can be rewritten as
the integral over the radial comoving distance, $r$,
\begin{equation}
\label{eq1b}
\Delta T ({\bf \hat{n}})
=  \frac{2}{c^3}\bar{T_0}\int_{0}^{r_{\rm LS}}
\dot{\Phi}(r{\bf \hat{n}})\, a \, dr,
\end{equation}
where $r_{\rm LS}$ is the radial comoving distance to the LS surface,
and $a$ is the scale factor of the Universe. 

Analogously, the integral for the projected CMB lensing potential due to scalar perturbations with no
anisotropic stress reads
\begin{equation}
\label{eq2b}
\phi({\bf \hat{n}})\equiv 
-2\int_0^{r_{\rm LS}} \frac{r_{\rm LS}-r}{r_{\rm LS}r}\,\frac{\Phi(r{\bf\hat{n}};c\eta_0-r )}{c^2}\,{\rm d}r\,,
\end{equation}
where $\eta_0$ is the present conformal time, and $\Phi$ is the physical
peculiar gravitational potential generated by density perturbations. For the purposes of this
work, the line-of-sight integration is made in the so-called
``Born-approximation'' along the \emph{undeflected} photon path,
which, for a given particle mass resolution, results to be accurate at
sub-percent level for weak-lensing and CMB-lensing calculations up to
$l\lesssim 3000$~\cite{Hilbert_etal_2007,Calabrese_etal_2015}. 

The corresponding deflection-angle integral is
\begin{equation}
\label{deflection_angle}
\boldsymbol{\alpha}({\bf \hat{n}})\equiv 
-2\int_0^{r_*}
\frac{r_{\rm LS}-r}{r_{\rm LS}r}\,\nabla_{\bf\hat{n}}\frac{\Phi(r{\bf\hat{n}};c\eta_0-r )}{c^2}\, {\rm d}r\,,
\end{equation}
where $[1/r]\nabla_{\hat{\bf n}}$ is
the two dimensional (2D) transverse derivative with respect to the
line-of-sight pointing in the direction ${\hat{\bf
n}}\equiv(\vartheta,\varphi)$.

We implement Eqs.~(\ref{eq1b})-(\ref{eq2b}) in our code
for CMB ray-tracing across the simulated $\Phi$ and $\dot{\Phi}$
distributions, in  order to produce all-sky ISW-RS and weak-lensing
maps, as described in \S~\ref{map-macking} below. 
\begin{figure*}[!ht]
\begin{center}
\setlength{\tabcolsep}{0.01pt}
\begin{tabular}{c c}
\includegraphics[angle=90,origin=c,width=0.5\textwidth]{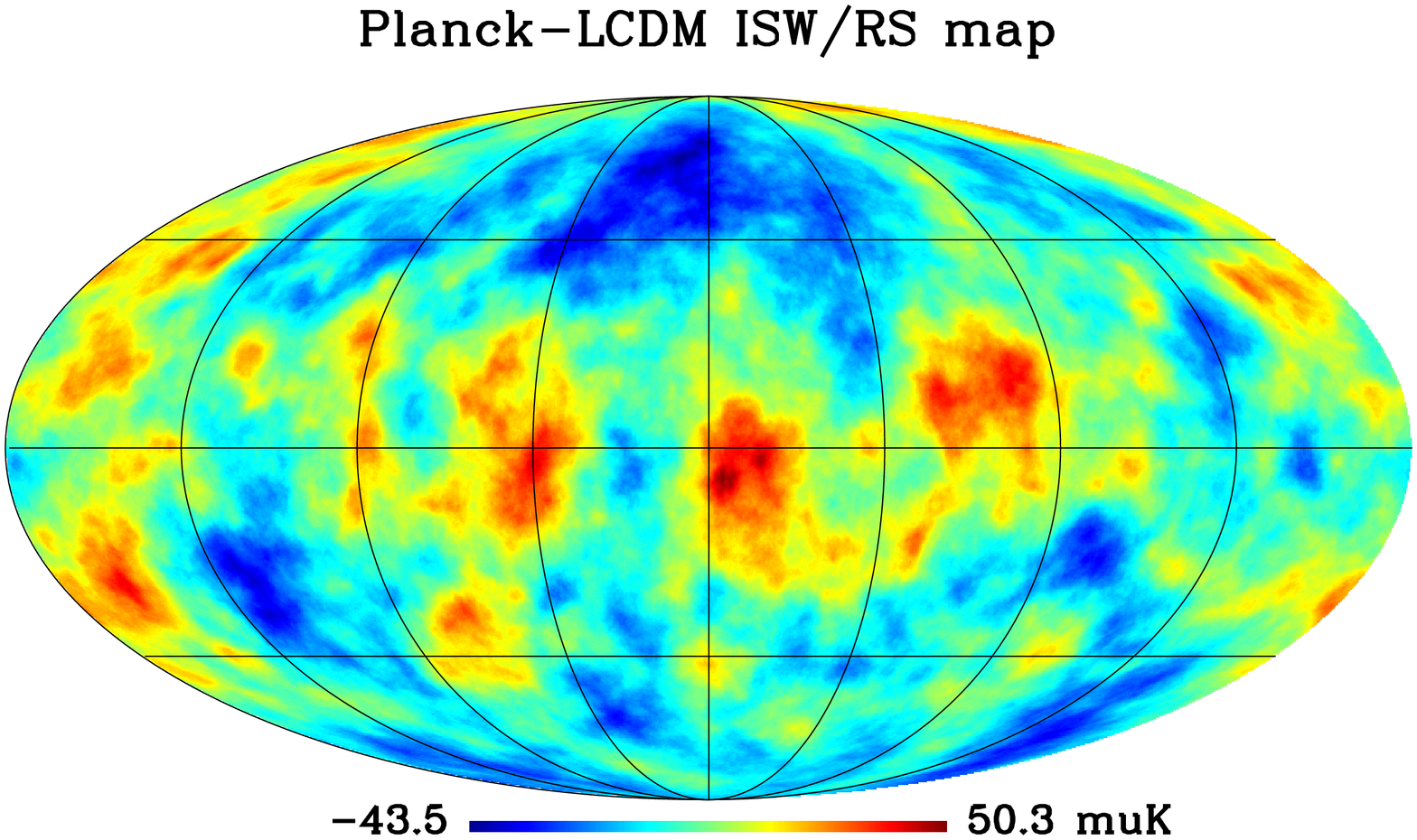}&
\includegraphics[angle=90,origin=c,width=0.5\textwidth]{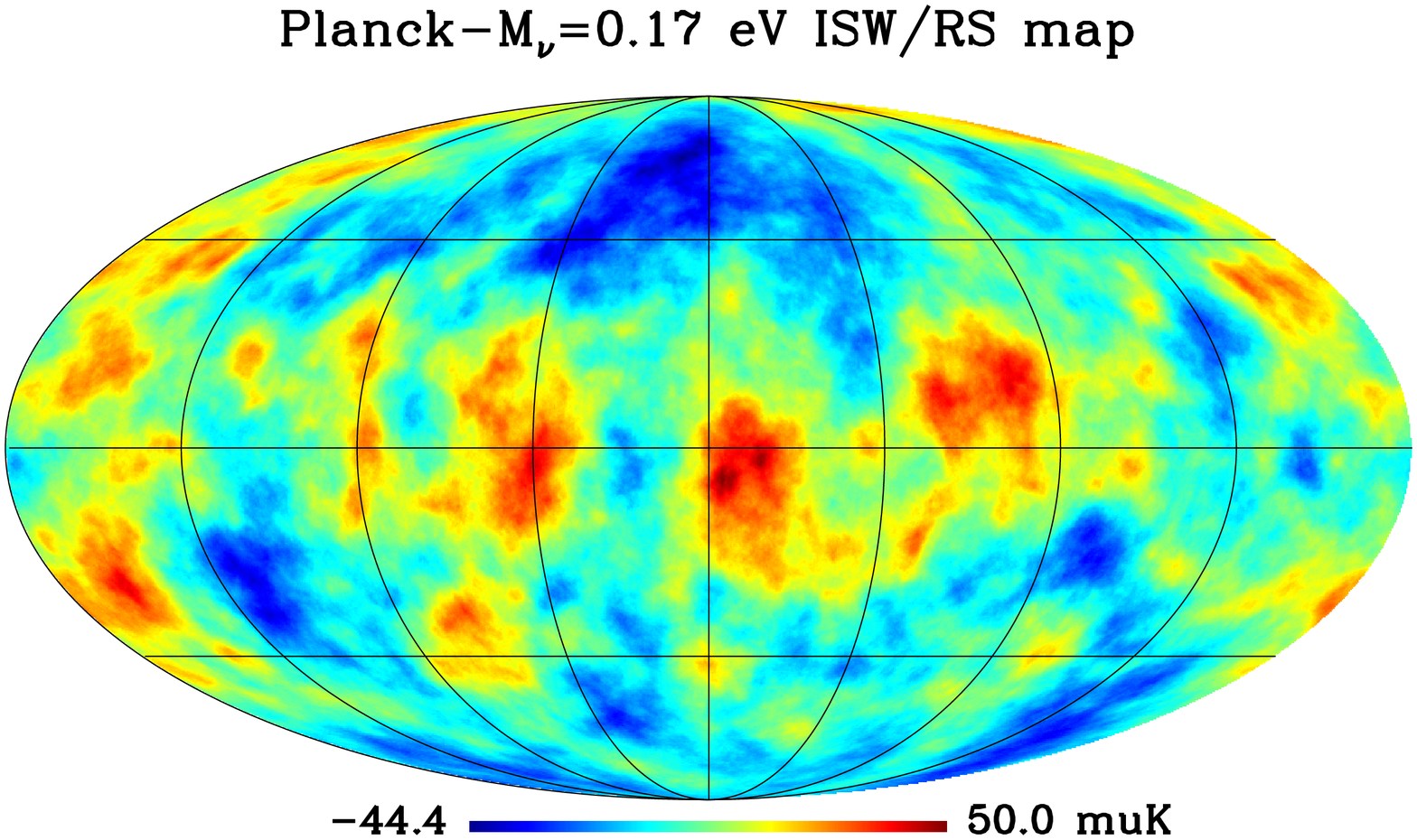}\\
\includegraphics[angle=90,origin=c,width=0.5\textwidth]{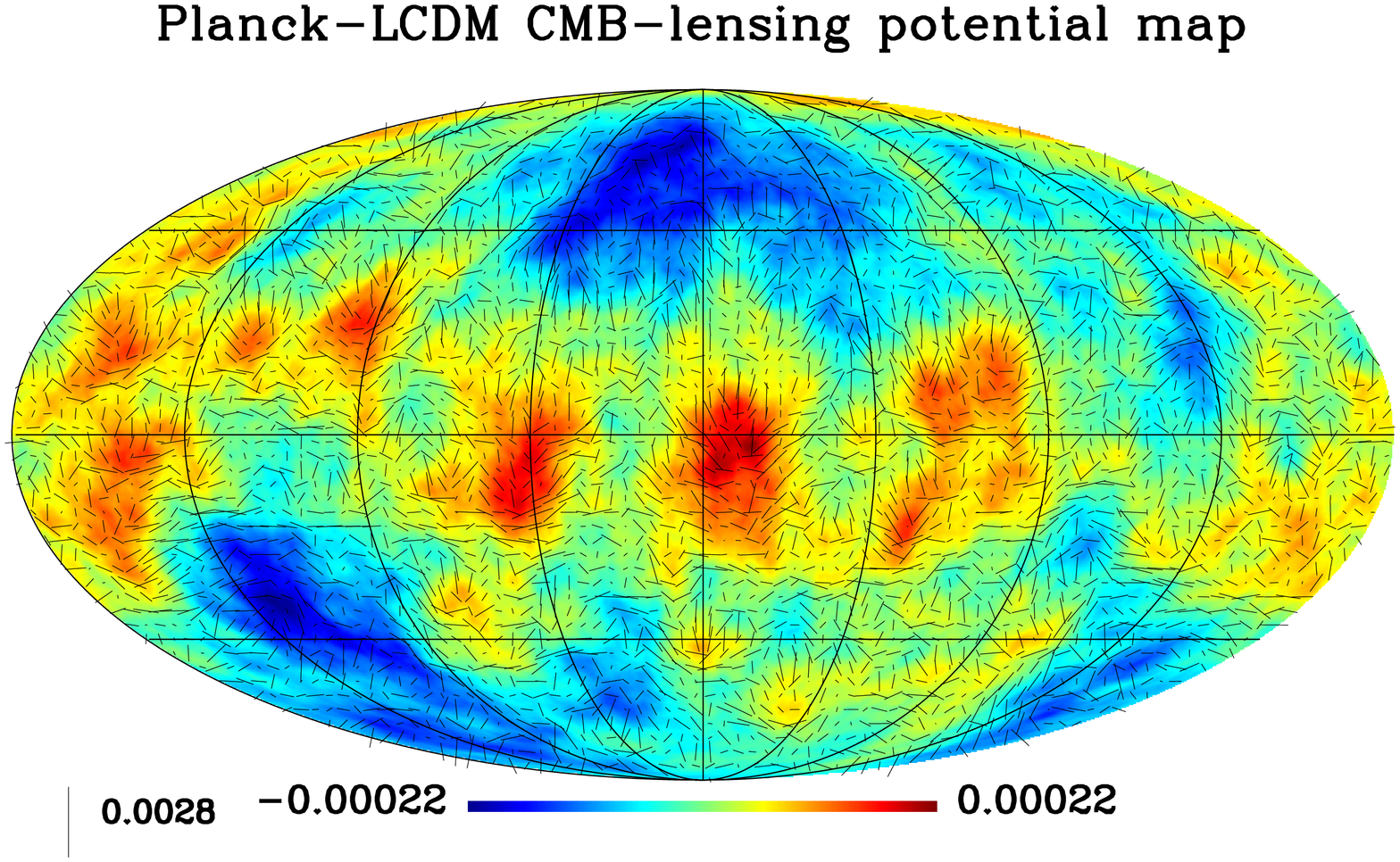}&
\includegraphics[angle=90,origin=c,width=0.5\textwidth]{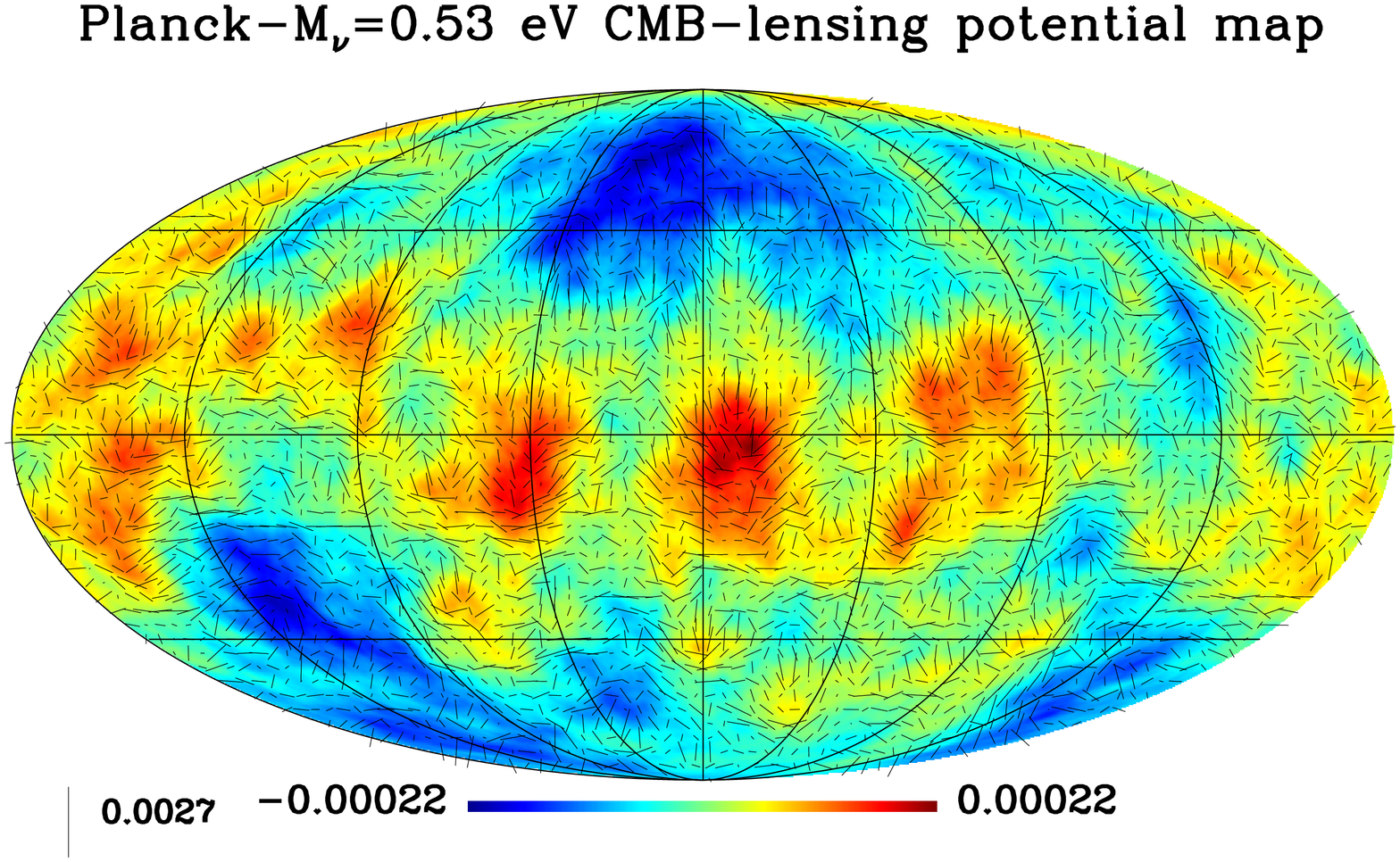}
 \end{tabular}
\end{center} 
\caption{Top panel: projected mono-dipole-subtracted full-sky maps of
the ISW-RS signal for $M_\nu=0$ eV (left) and $M_\nu=0.17$ eV
(right). They do not include primary
CMB anisotropies. 
Bottom panel: projected mono-dipole-subtracted full-sky
maps of the lensing potential for $M_\nu=0$ eV (left) and
$M_\nu=0.53$ eV (right); bars indicate the modulus and orientation of the
deflection-angle field $\boldsymbol{\alpha}$.}
\label{fig_maps}
\end{figure*}

\subsection{Map-making procedure}
\label{map-macking}
As mentioned in \S~\ref{demnuni}, during the production stage of the
DEMNUni simulations, using a properly modified version of
\gadgetthree, we have dumped on-the-fly 62 cubic grids of
$\Phi$ and $\dot{\Phi}$, with a mesh of $4096^3$ cells, each of $\simeq 0.5h^{-1}$Mpc on
a side\footnote{The gravitational potential itself has been 
calculated by first assigning the particles to the mesh with the
clouds-in-cells mass assignment scheme. The resulting density field has
then been Fourier transformed, multiplied with the Green's function of
the Poisson equation in Fourier space, and then transformed back to
obtain the potential. Also, a slight Gaussian smoothing on a scale $r_s$
equal to 1.25 times the mesh size has been applied in Fourier space in
order to eliminate residual anisotropies on the scale of the mesh, and a
deconvolution to filter out the clouds-in-cells mass assignment kernel
has been applied as well. The time derivative of the gravitational
potential has been computed via two-sided differentiation of two
potential grids dumped at two different step-times immediately
subsequent to the output times of each particle snapshot.}. Hence, the field
$\Phi$ corresponds to the density field of the simulations smoothed on
a scale of about $500\,h^{-1}{\rm kpc}$. This resolution is good enough to resolve scales $\gtrsim
10~h^{-1}$~Mpc, relevant for the effects analysed in this work.
As explained in \S~\ref{demnuni}, the resolution of the N-body
simulation (which contains structures down to the gravitational
softening length of $20\,h^{-1}{\rm kpc}$) is much greater, but not
necessary for the present study. 

In order to build mock all-sky maps of the CMB temperature anisotropies $\Delta T$
described in Eq.~(\ref{eq1}), we employ the map-making procedure
developed by~\cite{Carbone_etal_2007}, adapted to the ISW-RS effect,
\ie CMB photons are ray-traced along the undeflected line of sight through
the 3D field $\dot{\Phi}$. We apply the same kind
of ray-tracing also to the 3D $\Phi$-grids, in 
order to produce the same realisation of the Universe and compute the
cross-correlation signal between the ISW-RS temperature maps and the
CMB/weak-lensing potential maps.

To this aim, we stack the $\Phi$- and $\dot{\Phi}$-grids around the
observer, located at $z=0$, applying the replication and randomisation
procedure designed by~\cite{Carbone_etal_2007}.
This particular 3D tessellation scheme is required to avoid both the
repetition of the same structures along the line
of sight, and the generation of artifacts like
ripples in the simulated deflection-angle field, which can be avoided
only if the peculiar gravitational potential is continuous transversely to each
line of sight. With this procedure we produce a simulated volume
around the observer which is large
enough to carry out the integration over all the redshifts relevant
to this work. Finally, we select a pixelisation of the sky with a set of
directions ${\bf\hat{n}}\equiv(\vartheta,\varphi)$, following
the standard approach introduced by the
\textsf{HEALPix}\footnote{http://healpix.sourceforge.net} hierarchical
tessellation of the unit sphere~\cite{Healpix}.

In order to extract the impact of massive neutrinos on the ISW-RS
effect, mainly dominated by the free-streaming at high
redshifts~\cite{Lesgourgues_etal_2008}, we integrate $\dot{\Phi}$ along the line of sight up to $z
\simeq 21$ (for further details on the interpolation and integration
scheme see~\cite{Carbone_etal_2007}). For the simulated CMB lensing signal, previous
studies~\cite{Carbone_etal_2008,Carbone_etal_2012} indicate that an
integration up to $z \simeq 99$ is sufficient to recover mostly $\sim 99$\%
of the power. In the case of weak-lensing, we produce all-sky maps of
the lensing potential with different source redshifts $z_s$. In particular, we consider
that all the sources are placed on a spherical surface at redshifts
$z_s=2,5.5,8$, respectively. This is done for illustrative purposes,
\ie to quantify the impact of neutrino free-streaming at different
redshifts on the weak-lensing (WL) signal and its cross-correlation
with the total ISW-RS effect.

In the upper panel of Fig.~\ref{fig_maps}, we show projected
mono-dipole-subtracted full-sky maps of
the ISW-RS signal for the massless case (left top panel) and for $M_\nu=0.17$ eV (right top
panel). These maps represent the ISW-RS contribution alone to the CMB
temperature, Eq.~(\ref{eq1b}), and do not include primary CMB
anisotropies. In the lower panel, projected mono-dipole-subtracted full-sky
maps of the lensing potential
in the two cases, $M_\nu=0$ eV and $M_\nu=0.53$ eV, are
presented; bars here indicate the modulus and orientation of the
deflection angle field which represents the spatial gradient of the
lensing potential map, as defined in Eq.~(\ref{deflection_angle}).
These maps have been obtained via the
map-making technique described above, using a \textsf{HEALPix} pixelisation
parameter $N_{\rm side}=2048$, and have an angular resolution of $\simeq
1.72^\prime$, with 50331648 pixels in total.
In Fig.~\ref{fig_maps}, the correlation between the maps in the upper
and lower panels, \ie between the ISW-RS and lensing potential
realisations, is clearly visible, and shows how the origin of these two effects is due to
the same LSS distribution crossed by CMB photons.

\begin{figure*}[!ht]
\begin{center}
\setlength{\tabcolsep}{0.01pt}
\begin{tabular}{c c}
\includegraphics[angle=90,origin=c,width=0.5\textwidth]{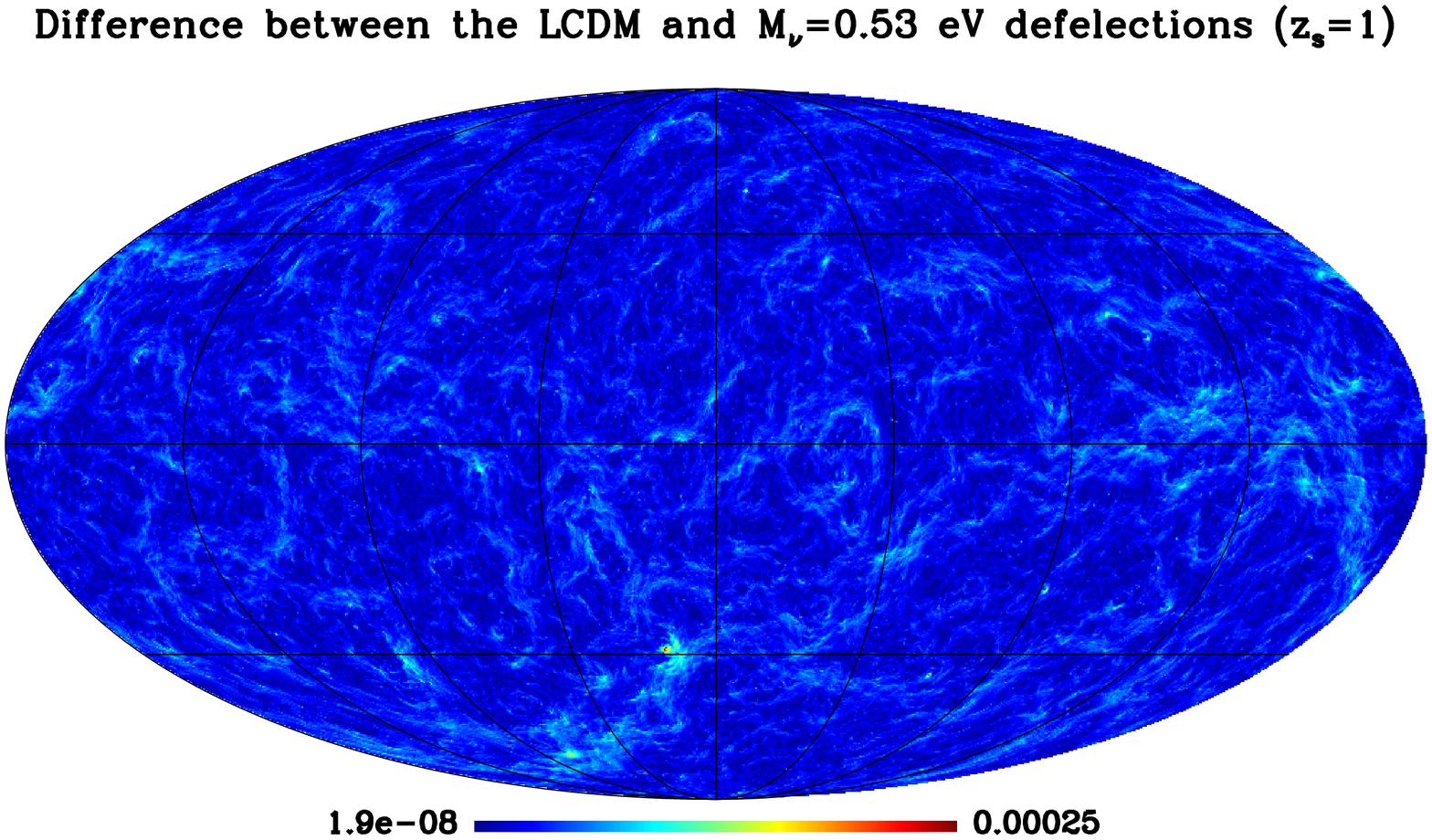}&
\includegraphics[angle=90,origin=c,width=0.5\textwidth]{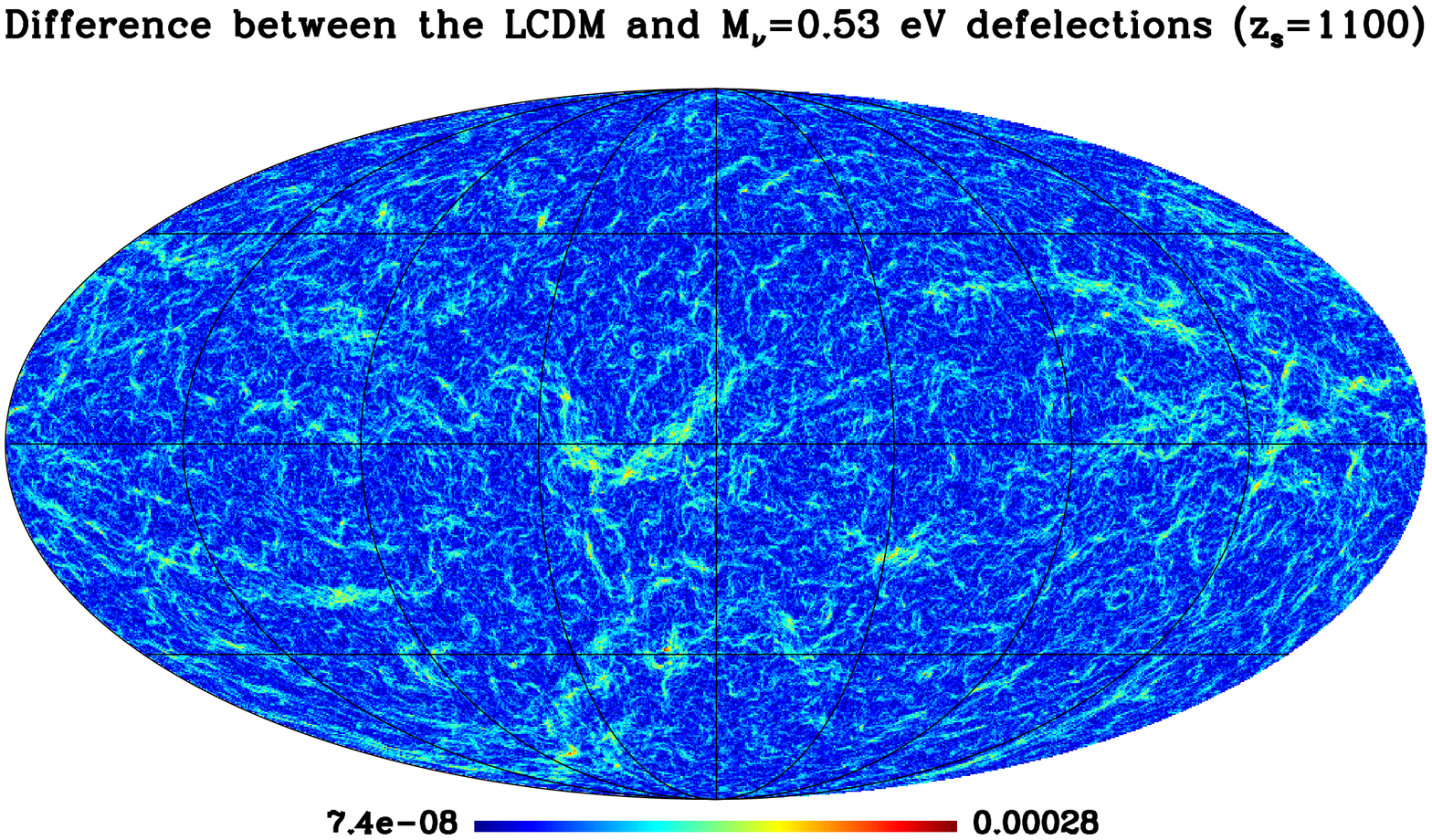}\\
 \end{tabular}
\end{center} 
\caption{Modulus of the deflection-angle vector difference between the massless case, $M_\nu=0$ eV, and the
massive neutrino case, $M_\nu=0.53$ eV. Left: all the
galaxy sources are supposed to be located at $z_s=1$. Right: the source is represented by CMB photons at
$z_s=1100$.}
\label{fig_diff_maps}
\end{figure*}

\section{The impact of massive neutrinos on auto power spectra}
\label{auto-power}
Massive neutrinos produce noticeable effects on CMB secondary
anisotropies, which in some cases, \eg for large neutrino masses,
can be even visually inspected. Let us first notice some
visible differences between the massless and massive neutrino cases:
in the lower panels of Fig.~\ref{fig_maps}, the neutrino
free-streaming suppresses, as expected, the lensing potential signal
with respect to the $\Lambda$CDM scenario (this can be especially observed by
an eye inspection of the filaments in the two maps, and from the bar
units of the modulus of the deflection angle); on the other
hand, in the upper panels, the effect is opposite, \ie in the 
ISW-RS case the neutrino free-streaming produces a slight excess of
power with respect to the massless case, and this is caused by the
$\dot{\Phi}$ induced by hot neutrinos, mostly at high redshifts and
for lighter neutrino masses (for this reason the temperature range represented by the colour bar is
larger for the $M_\nu=0.17$ eV case). We
will explain this effect in more details in the next section, but let
us anticipate that smaller neutrino masses produce a
larger excess in the CMB temperature power spectrum.

In Fig.~\ref{fig_diff_maps} we show the modulus of the deflection
angle difference, between the massless case, $M_\nu=0$ eV, and the
massive neutrino case, $M_\nu=0.53$ eV.  Here the plotted quantity is
$\sqrt{(\Delta \alpha_1)^2+(\Delta
\alpha_2)^2}$, where $\alpha_1$ and $\alpha_2$ are the two components of the
deflection angle $\boldsymbol{\alpha}$, and $\Delta \alpha_i$, with $i=1,2$, stands for the
difference between each component in the two cases $M_\nu=0$ eV and
$M_\nu=0.53$ eV.
In the left panel all the
galaxy sources are supposed to be located at $z_s=1$, while in the
right panel the source is represented by CMB photons at
$z_s=1100$. These maps visually show the suppression in structure
formation due to the presence of massive neutrinos, for two different cases of
weak-lensing. Given the limited spatial resolution of the potential grids,
smaller angular scales are captured via ray-tracing as
the source redshift increases, such that the right panel in
Fig.~\ref{fig_diff_maps} shows much more ``non-linear'' features with respect to the left panel.

\begin{figure*}[!ht]
\begin{center}
\setlength{\tabcolsep}{0.01pt}
\begin{tabular}{c c}
\includegraphics[width=0.5\textwidth]{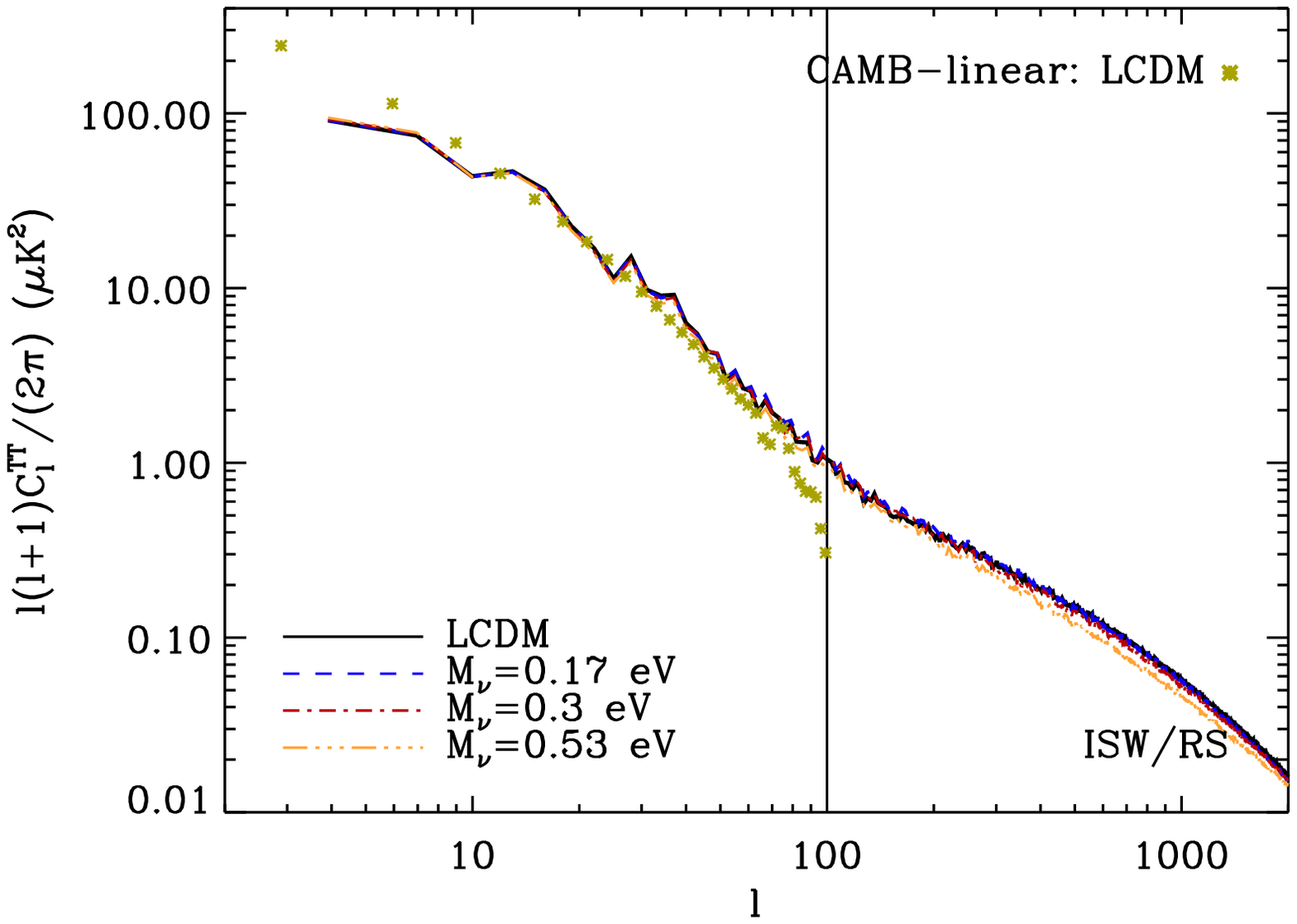}&
\includegraphics[width=0.5\textwidth]{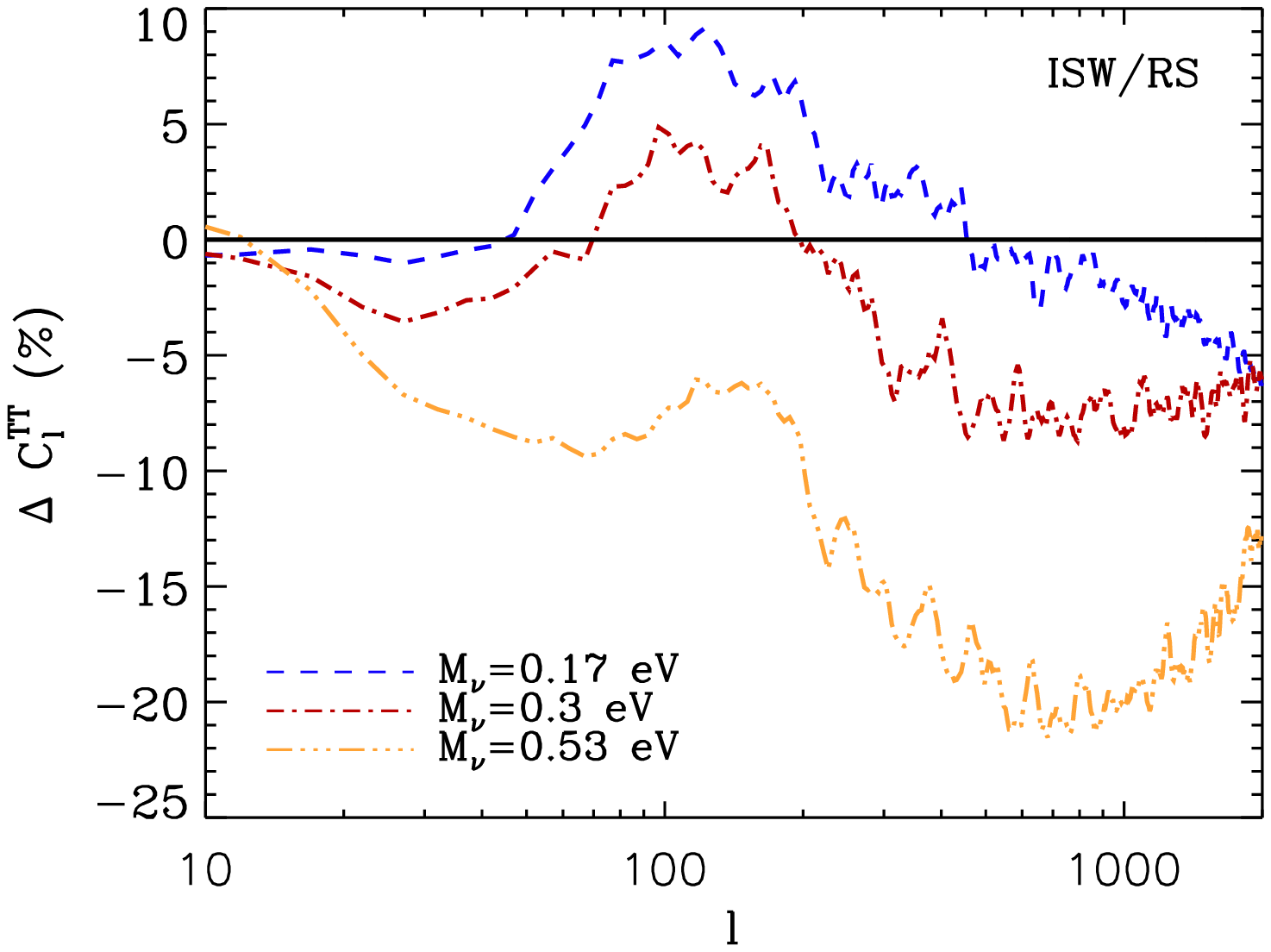}
 \end{tabular}
\end{center} 
\caption{Left: simulated angular power spectra of the total ISW-RS induced temperature
anisotropies, for $M_\nu=0,0.17,0.3,0.53$ eV (black solid, blue
dashed, red dot-dashed, and orange tri-dot-dashed lines,
respectively). Light-green symbols represent the linear contribution (ISW alone) from
\CAMB\ in the massless case, the vertical line at
$l\simeq 100$ corresponds roughly to the transition between the linear and
non-linear regimes. 
Right: percent residuals, with respect to the
massless case, of the  ISW-RS power spectra obtained via direct ray-tracing across the simulations.}
\label{fig_iswrs}
\end{figure*}

\begin{figure*}[!ht]
\begin{center}
\setlength{\tabcolsep}{0.01pt}
\begin{tabular}{c c} 
\includegraphics[width=0.5\textwidth]{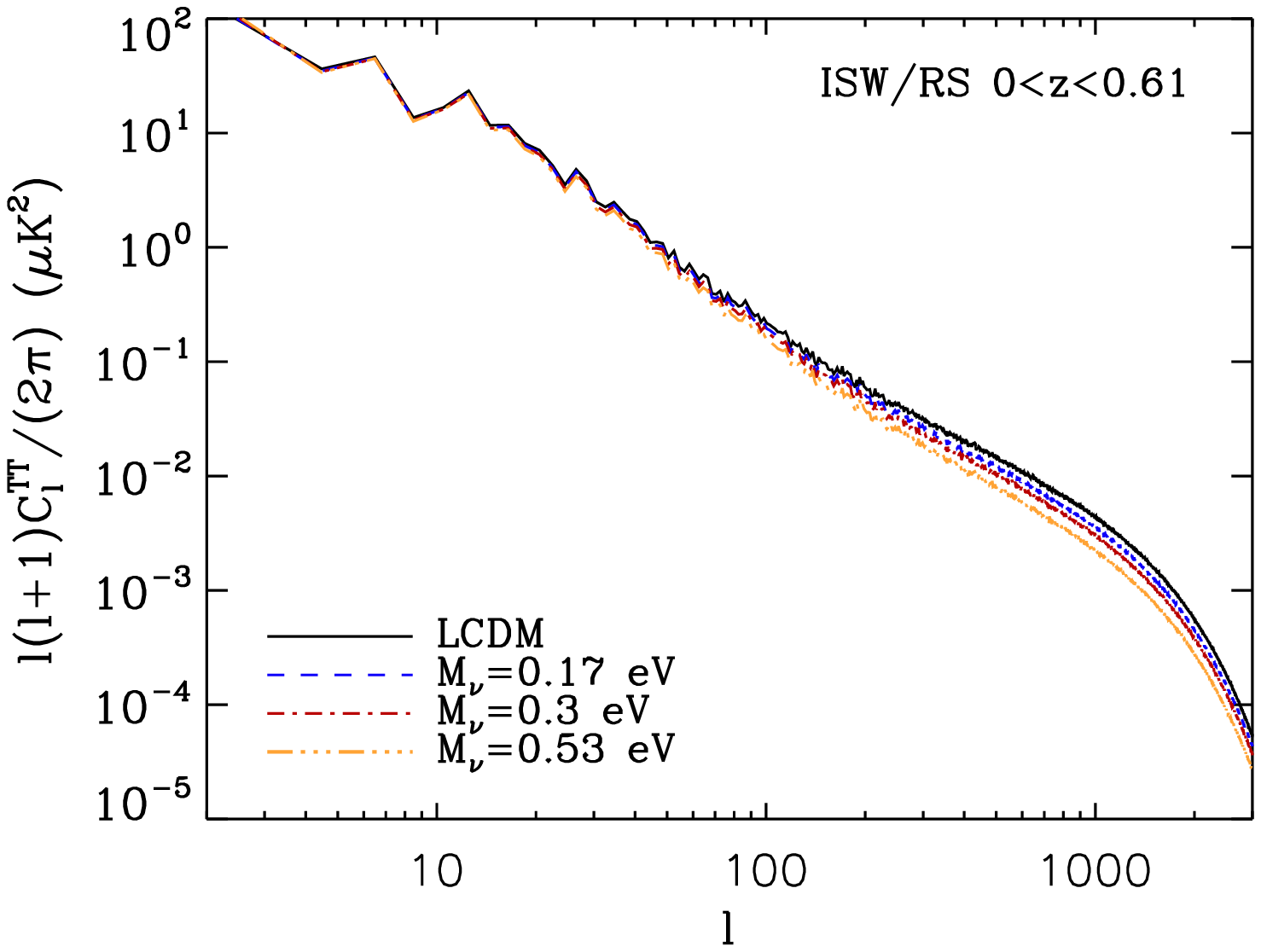}&
\includegraphics[width=0.5\textwidth]{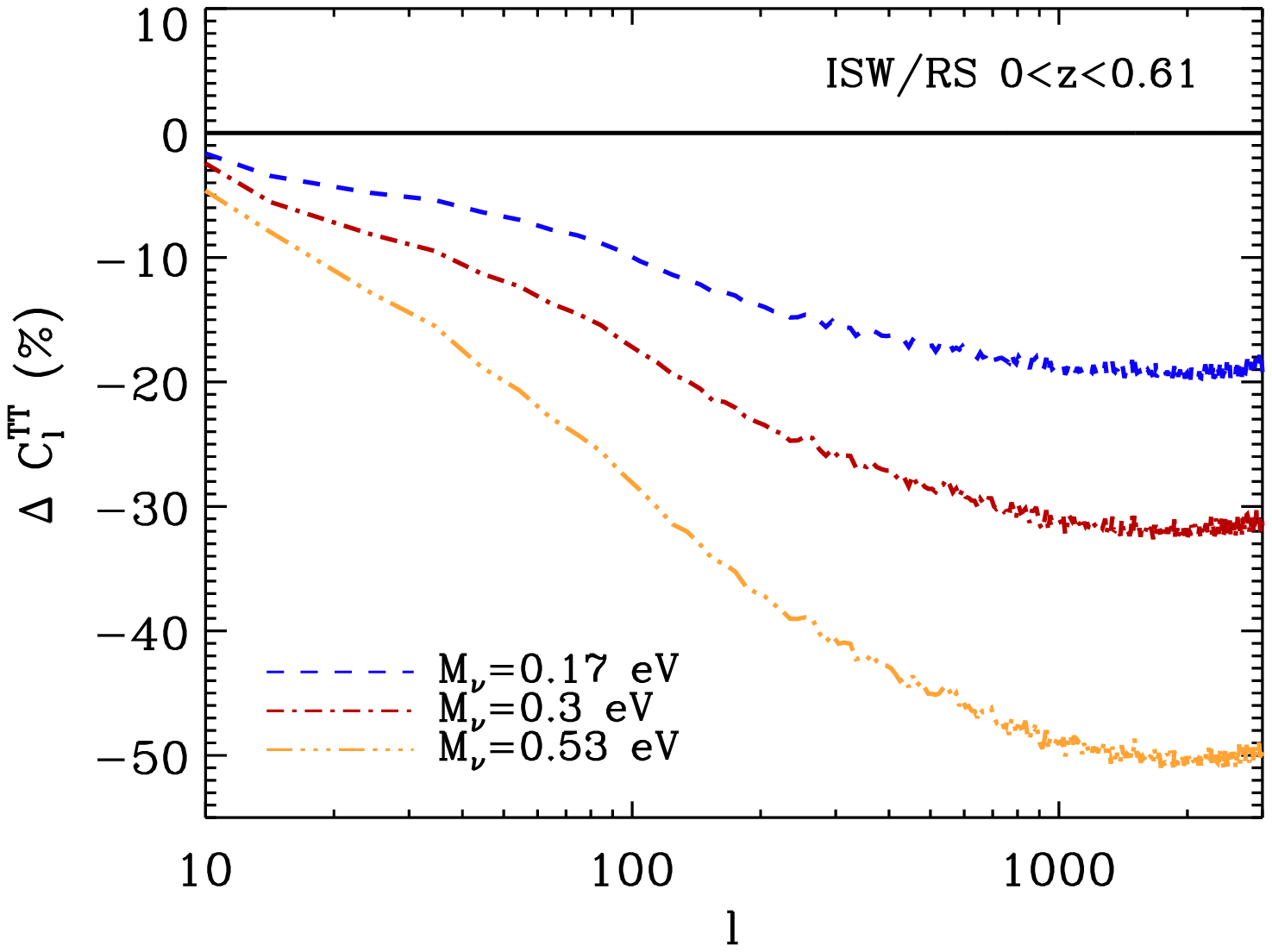}\\
\includegraphics[width=0.5\textwidth]{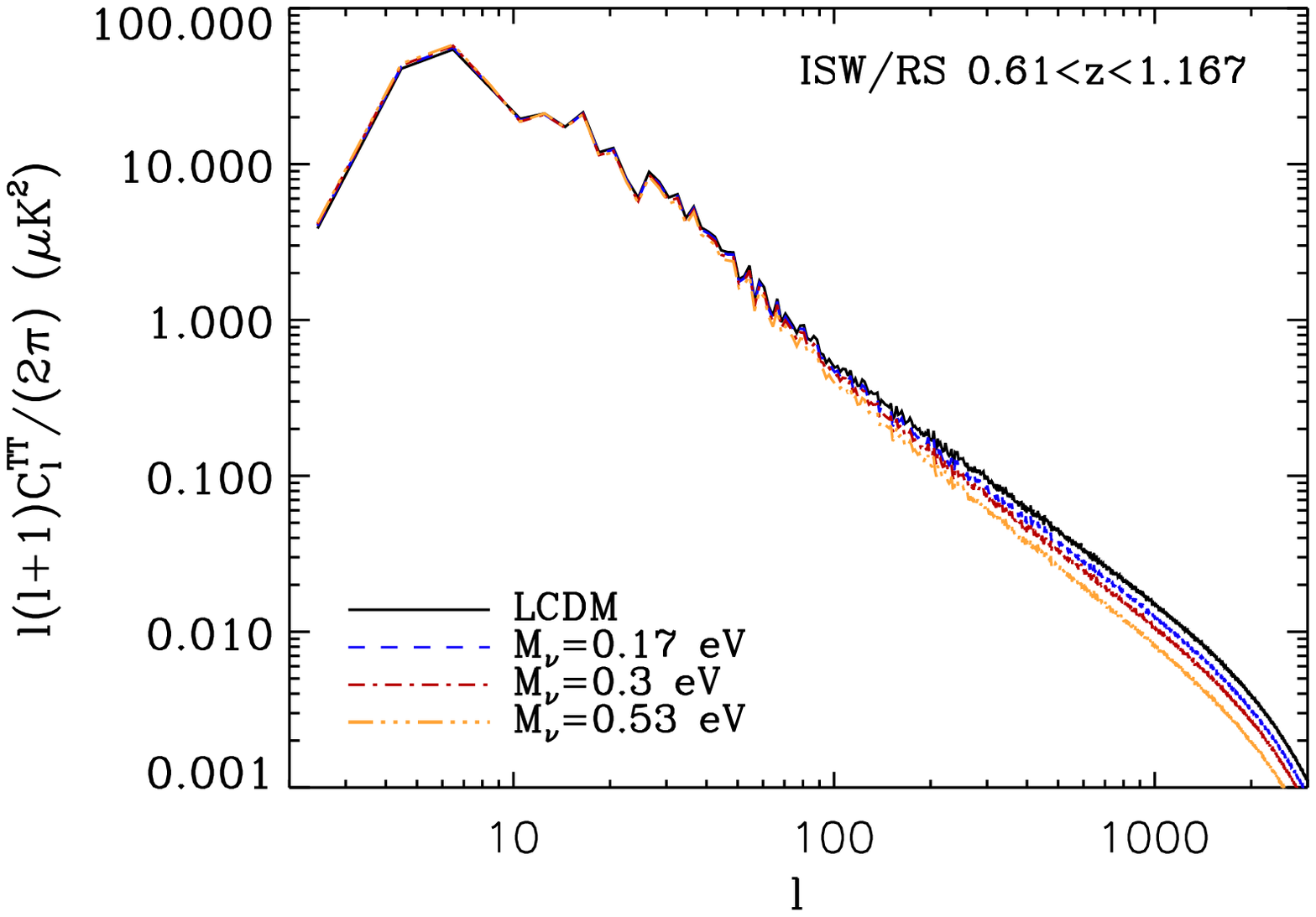}&
\includegraphics[width=0.5\textwidth]{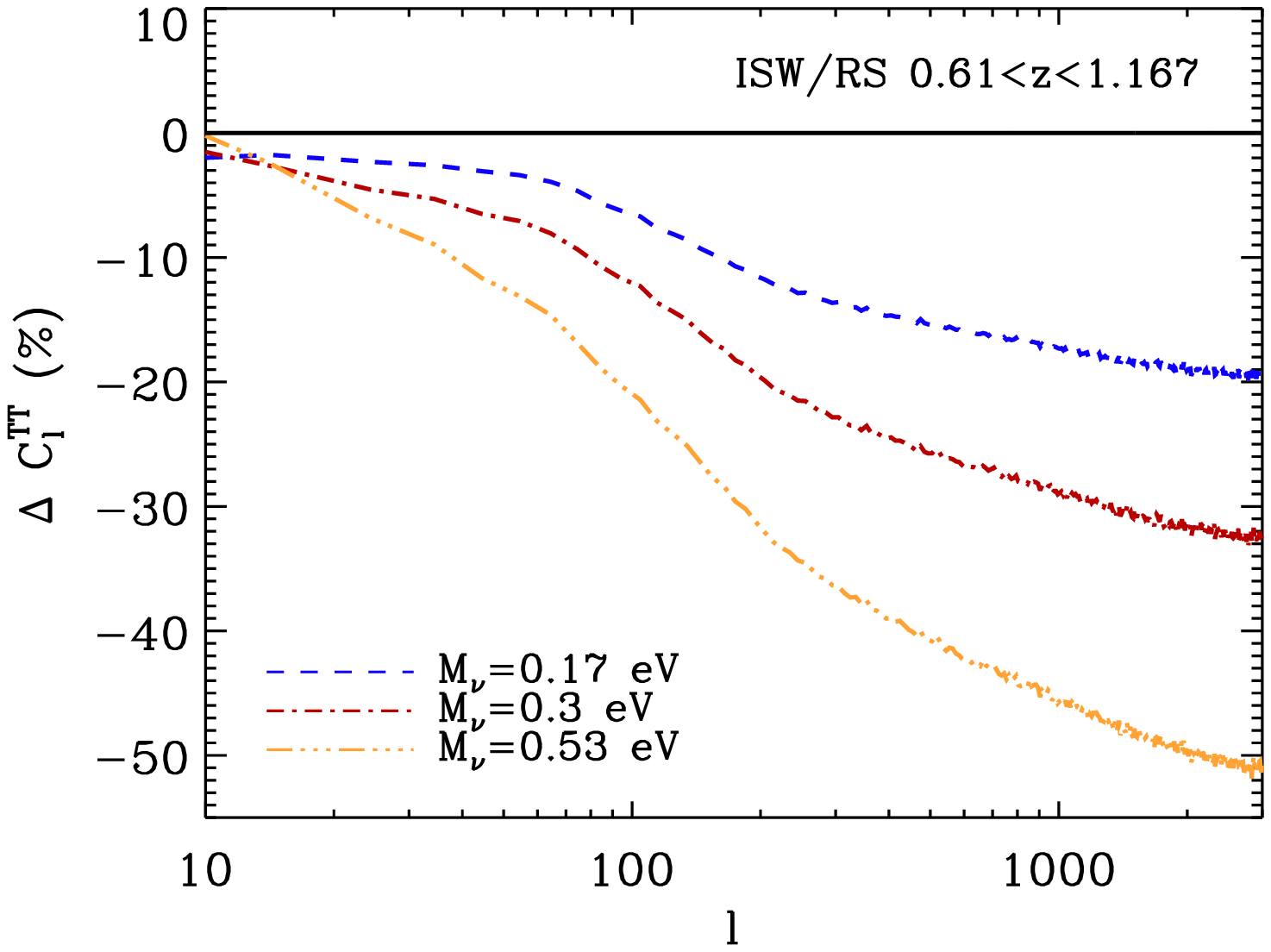}\\
\includegraphics[width=0.5\textwidth]{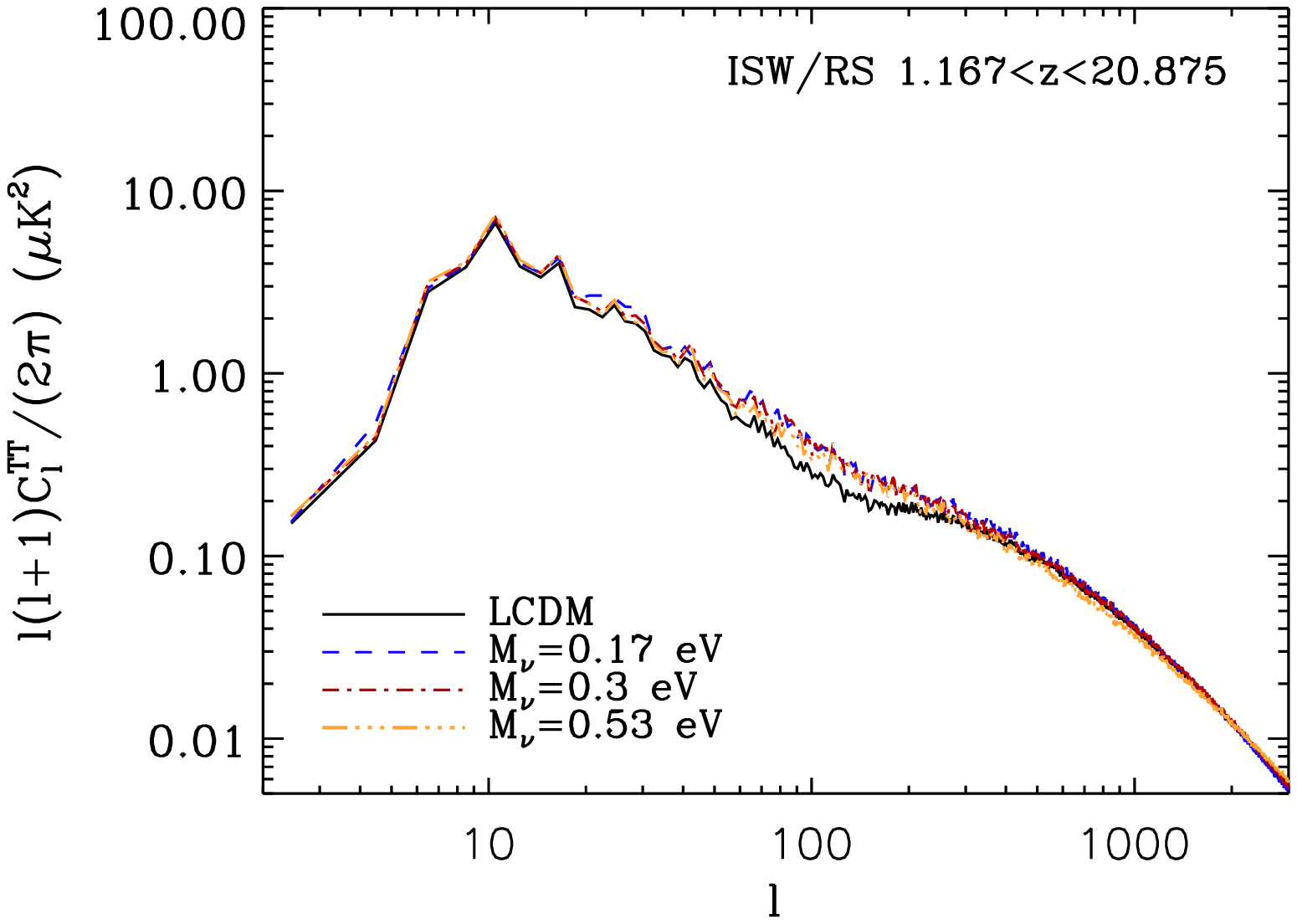}&
\includegraphics[width=0.5\textwidth]{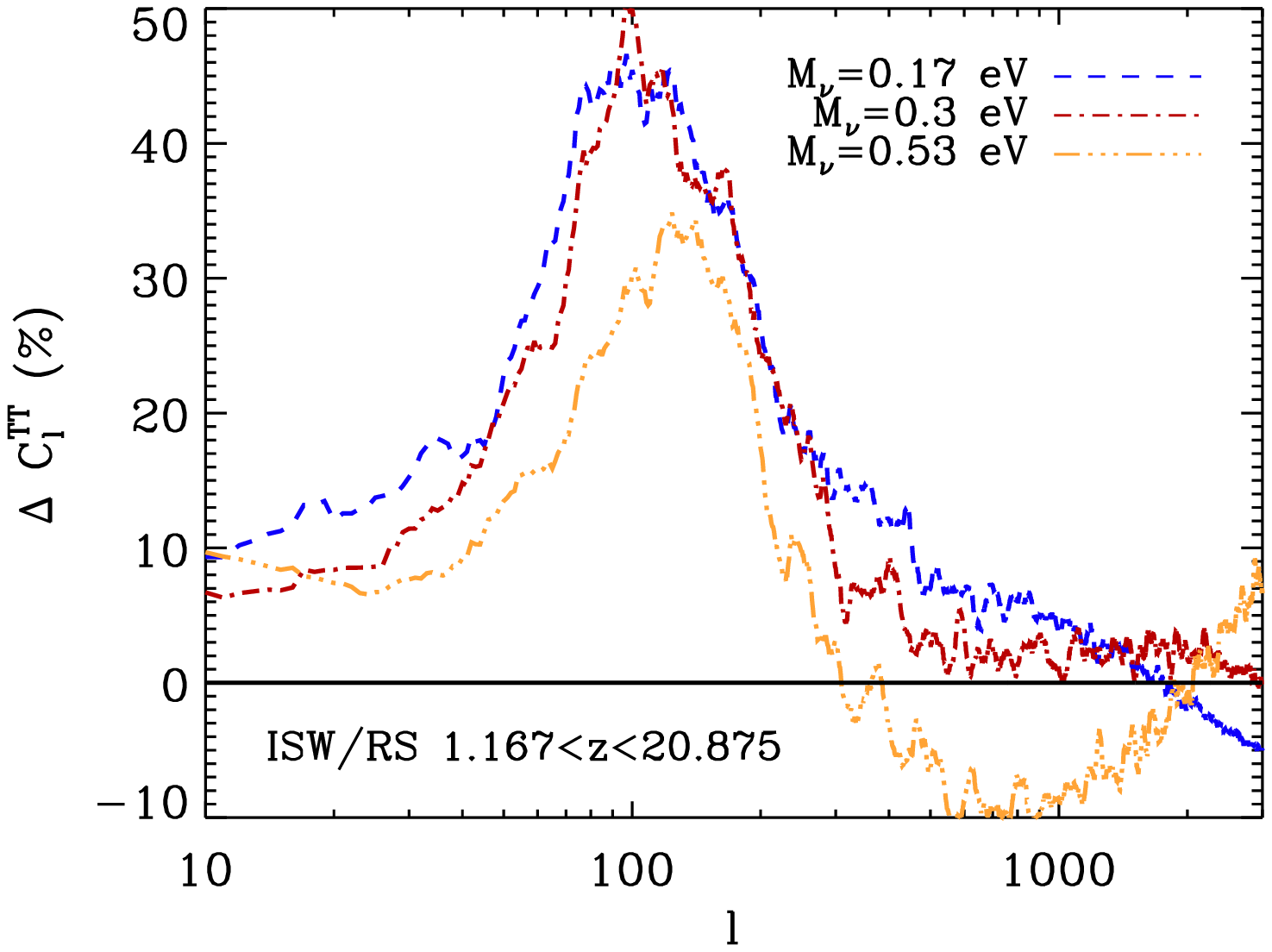}
\end{tabular}
\end{center} 
\caption{Left: simulated power spectra of the ISW-RS induced
anisotropies computed by integration in different redshift bins. Right: corresponding percent residuals with respect to the massless case.}
\label{fig_iswrs_z}
\end{figure*}

\subsection{The total ISW-RS effect}
\label{ISWRS}
The simulated power spectra of the total ISW-RS induced temperature
anisotropies, obtained via ray-tracing from $z=0$ to $z\simeq 21$, are
shown in the left panel of Fig.~\ref{fig_iswrs} for
four different total neutrino masses, $M_\nu=0,0.17,0.3,0.53$ eV. They
are represented by the black (solid), blue (dashed), red (dot-dashed),
and orange (tri-dot-dashed) lines, respectively. The light-green
symbols represent the linear contribution (ISW alone) from
\CAMB\footnote{http://camb.info} in the massless case, while the vertical line at
$l\simeq 100$ corresponds roughly to the transition between the linear and
non-linear regimes \cite{Scahefer2006,Scahefer2008}. Let us notice the good agreement between the
theoretical expectations from \CAMB\ and the simulation outputs at
multipoles $l>10$; at $l<10$ the simulated signals show a lack of power with
respect to the predictions, due to the finite size of the
simulation box; on the other hand we succeed in reproducing the non-linear
contribution for $l>100$, scales at which \CAMB\ predictions fail, 
given that the RS contribution is not implemented in the code.
In the right panel of Fig.~\ref{fig_iswrs} it is possible to appreciate
the differences produced in the ISW-RS induced temperature
anisotropies by free-streaming neutrinos with different total
masses. In particular, here we show the 
percent residuals with respect to the massless case; for
small masses, $M_\nu=0.17\,,0.3$ eV, we observe an excess of power of
about $10$\% and $5$\% respectively, at  $50 \lesssim l \lesssim
150$, corresponding to the transition  
from linear to non-linear regimes. This excess is indeed expected to
originate from the $\dot{\Phi}$ term induced by the slow decay of 
gravitational and matter perturbations produced by hot neutrinos at
intermediate cosmological scales, happening both during the matter and
dark energy domination eras. On the other hand, for large neutrino masses, \eg
$M_\nu=0.53$ eV, which correspond to smaller neutrino thermal
velocities, the total effect consists in a power suppression, similar to what
happens for the lensing potential in the presence of massive
neutrinos. 
Finally, on fully non-linear scales $l>200$, where the signal is
totally due to the RS effect, massive neutrinos decrease the total
temperature power with respect to the massless scenario, and this suppression increases
with the neutrino mass. This is reasonably due to the
non-linear excess of suppression of the total matter power spectra
with respect to linear expectations in the presence of massive
neutrinos (see the left panel of Fig.~\ref{pk} for $k\lesssim 10$ $h$/Mpc,
corresponding to the $\dot{\Phi}$-grid resolution), which for lower
thermal velocities may dominate the total effect.

In order to understand how the effect of free-streaming massive
neutrinos evolves with $z$, in the left panels of
Fig.~\ref{fig_iswrs_z} we show the power spectra of the ISW-RS induced
anisotropies computed in different redshift bins. We observe that for
low redshifts, $z\lesssim 1$, neutrinos produce a larger suppression in $C_l^{TT}$ at
$l>100$ as their mass increases; in this case the slow down of
structure formation in the presence of massive neutrinos has the
dominant role. On the other hand, for larger
redshifts, $z\gtrsim 1$, far from the epoch of recent acceleration, we
see an excess of power which peaks at the transition scale, $l\simeq
100$, and is larger for lighter neutrinos; as pointed out in
\cite{Lesgourgues_etal_2008}, this is due to the behaviour of the
linear growth rate in the presence of massive neutrinos, which
produces an evolving gravitational potential even in the absence of
dark energy. 

More quantitatively, in the right panels of
Fig.~\ref{fig_iswrs_z}, we show the percent residuals with respect to
the $\Lambda$CDM case: in the upper panels, for $z\lesssim 1$, the
power suppression in the $C_l^{TT}$ has a trend similar to the matter
$P(k)$, increasing at large multipoles and somewhat proportional to
the neutrino mass ratios. In the lower panel, instead, at higher redshifts,
$1\lesssim z \lesssim 21$, the produced excess of power can even reach
$50$\% for small neutrino masses at $l\simeq 100$. The combination of the lack
of power at low redshifts and the excess at high redshifts produces
the total effect shown in the left panel of Fig.~\ref{fig_iswrs}. 

Incidentally, let us notice how the impact of the limited simulation volume increases
with redshift, producing a larger lack of power at $l<10$ for larger $z$ values. 

\begin{figure*}[!ht]
\begin{center}
\setlength{\tabcolsep}{0.01pt}
\begin{tabular}{c c}
\includegraphics[width=0.5\textwidth]{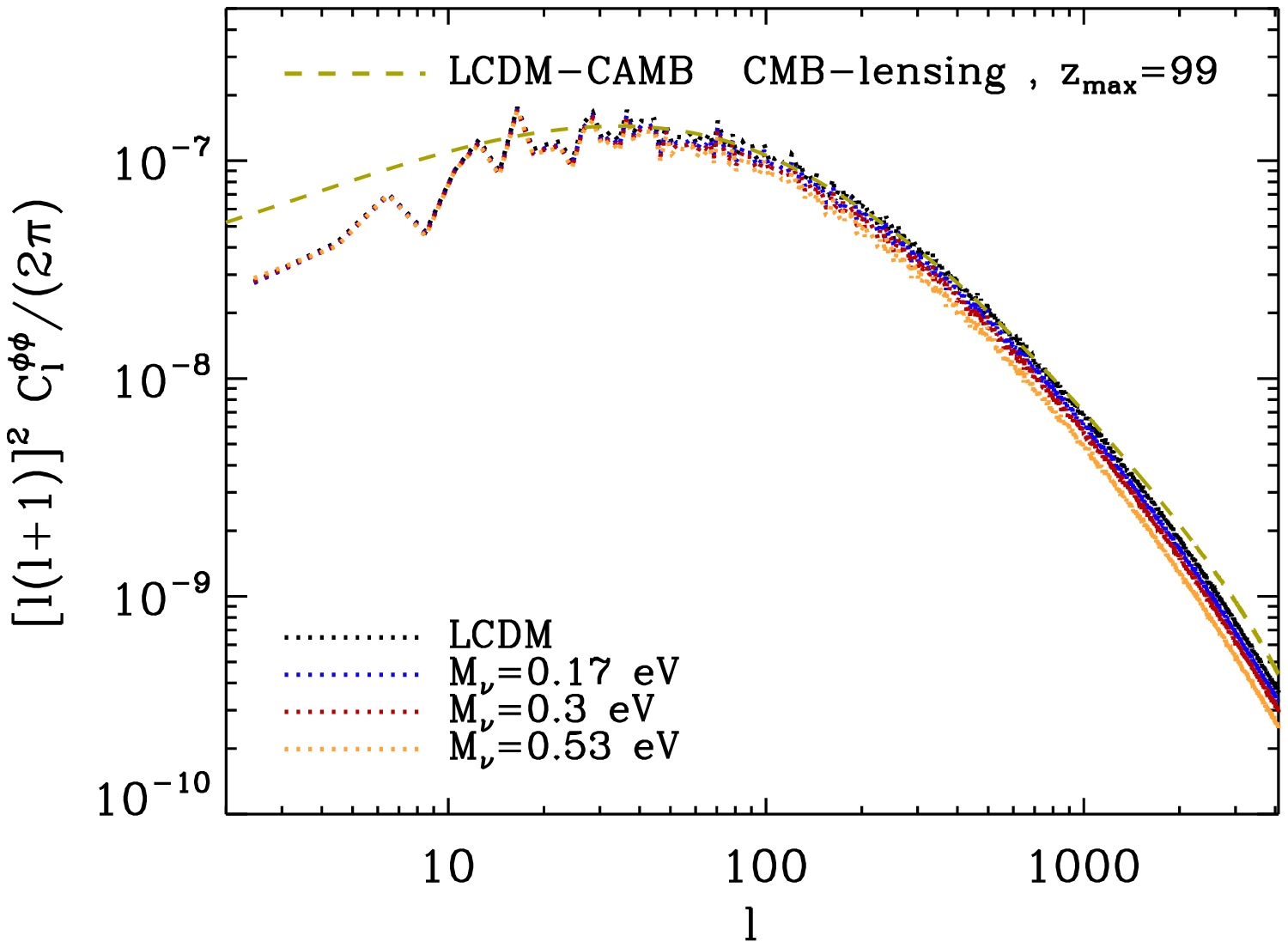}&
\includegraphics[width=0.5\textwidth]{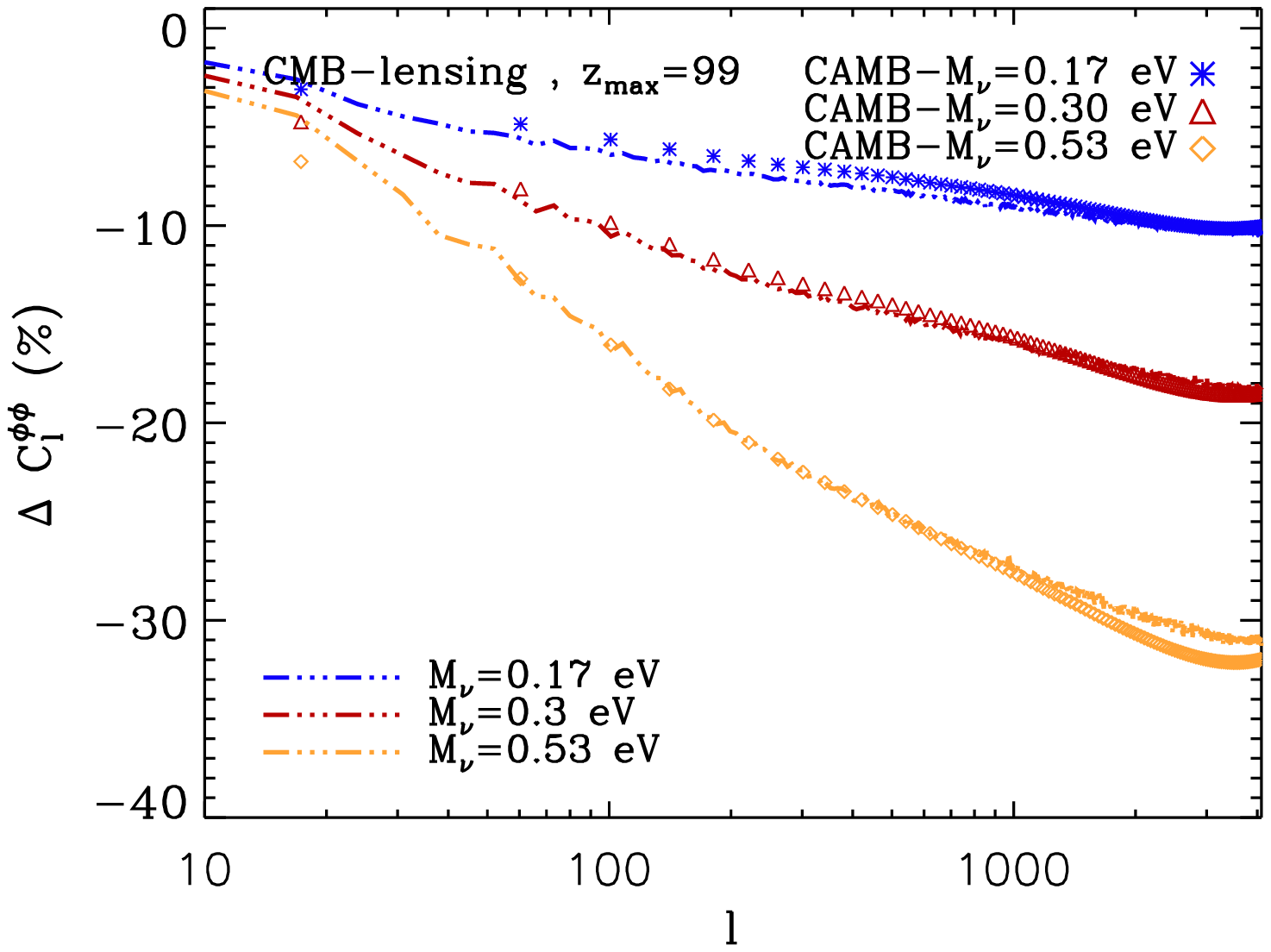}\\
\includegraphics[width=0.5\textwidth]{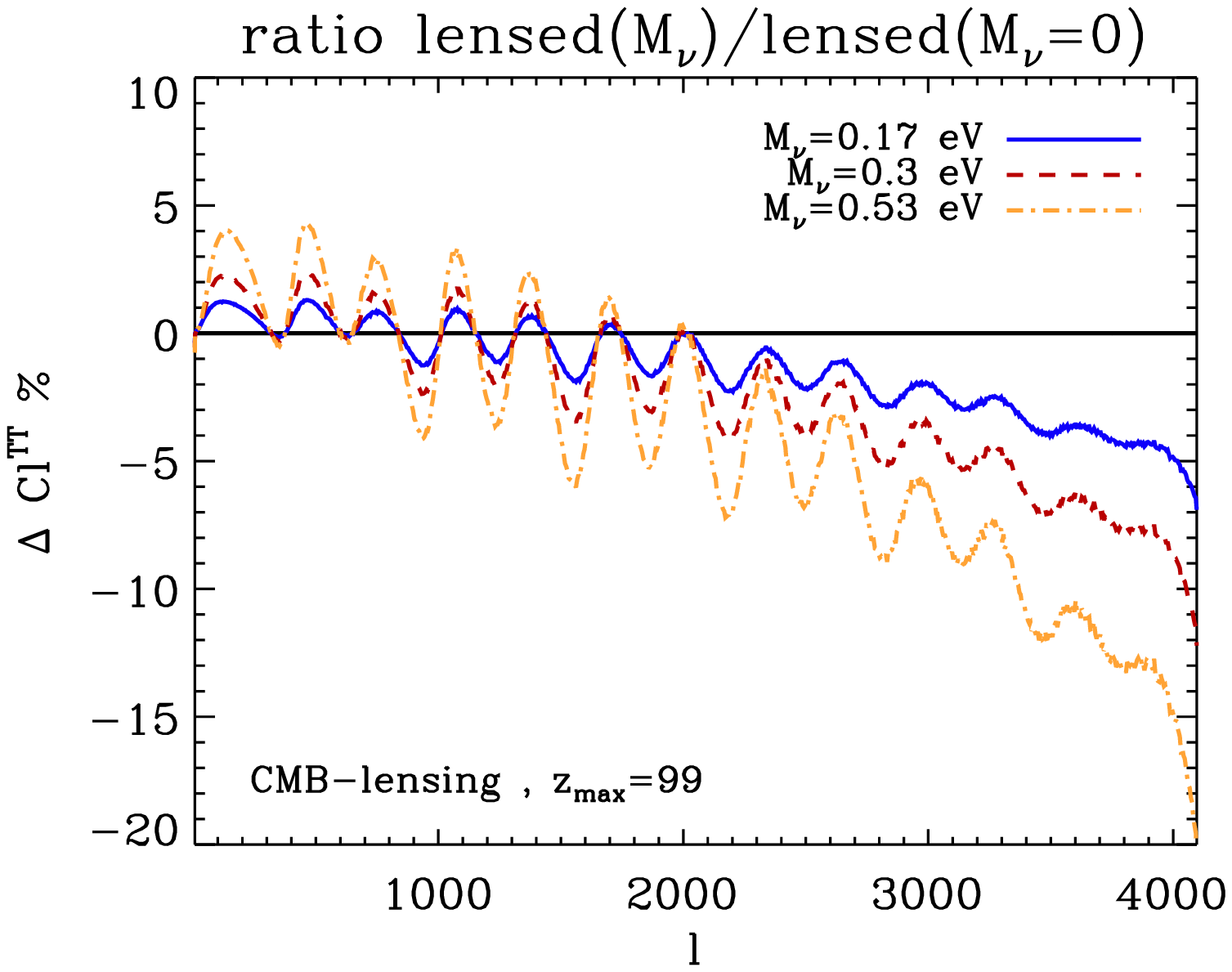}&
\includegraphics[width=0.5\textwidth]{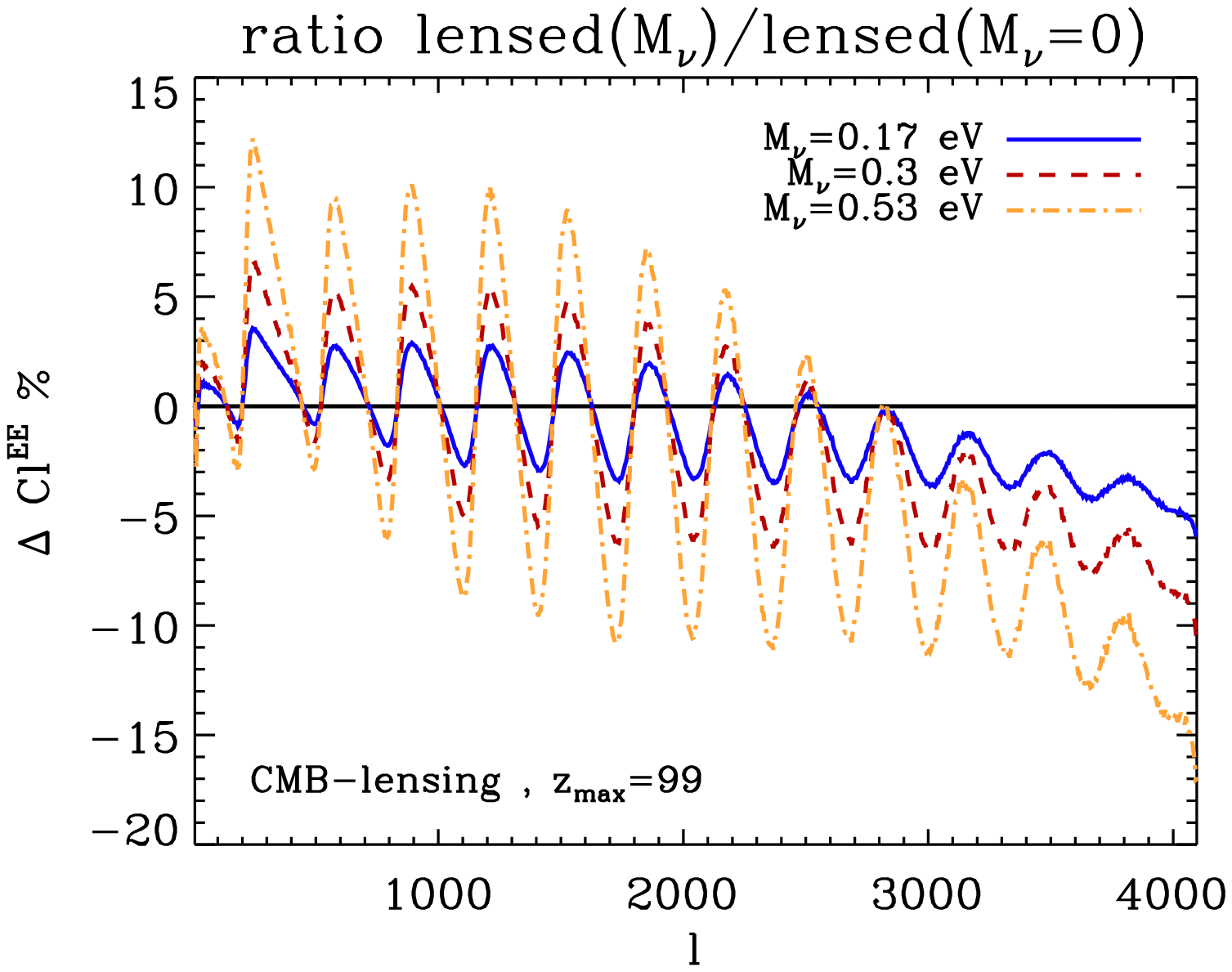}\\
\includegraphics[width=0.5\textwidth]{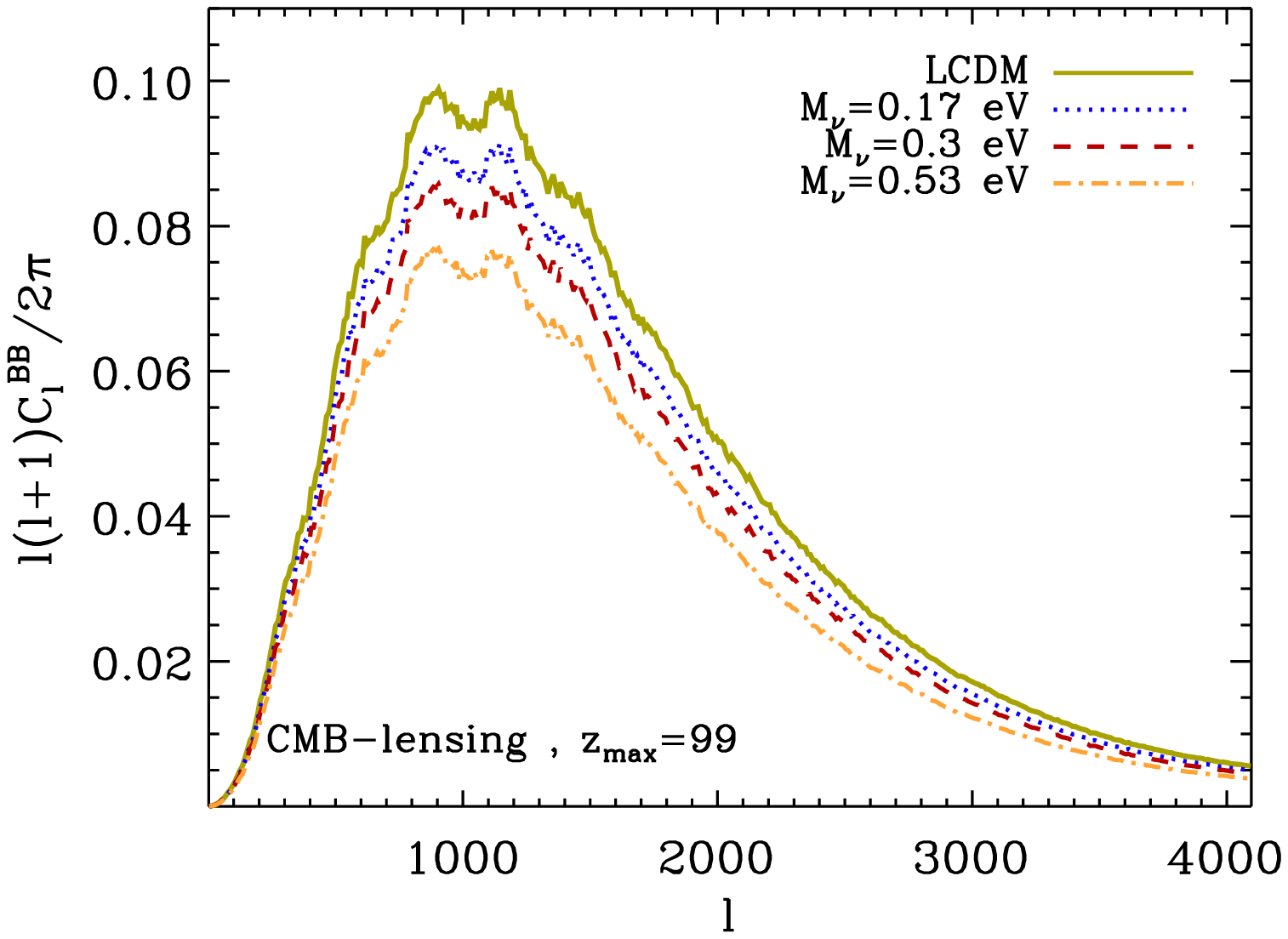}&
\includegraphics[width=0.5\textwidth]{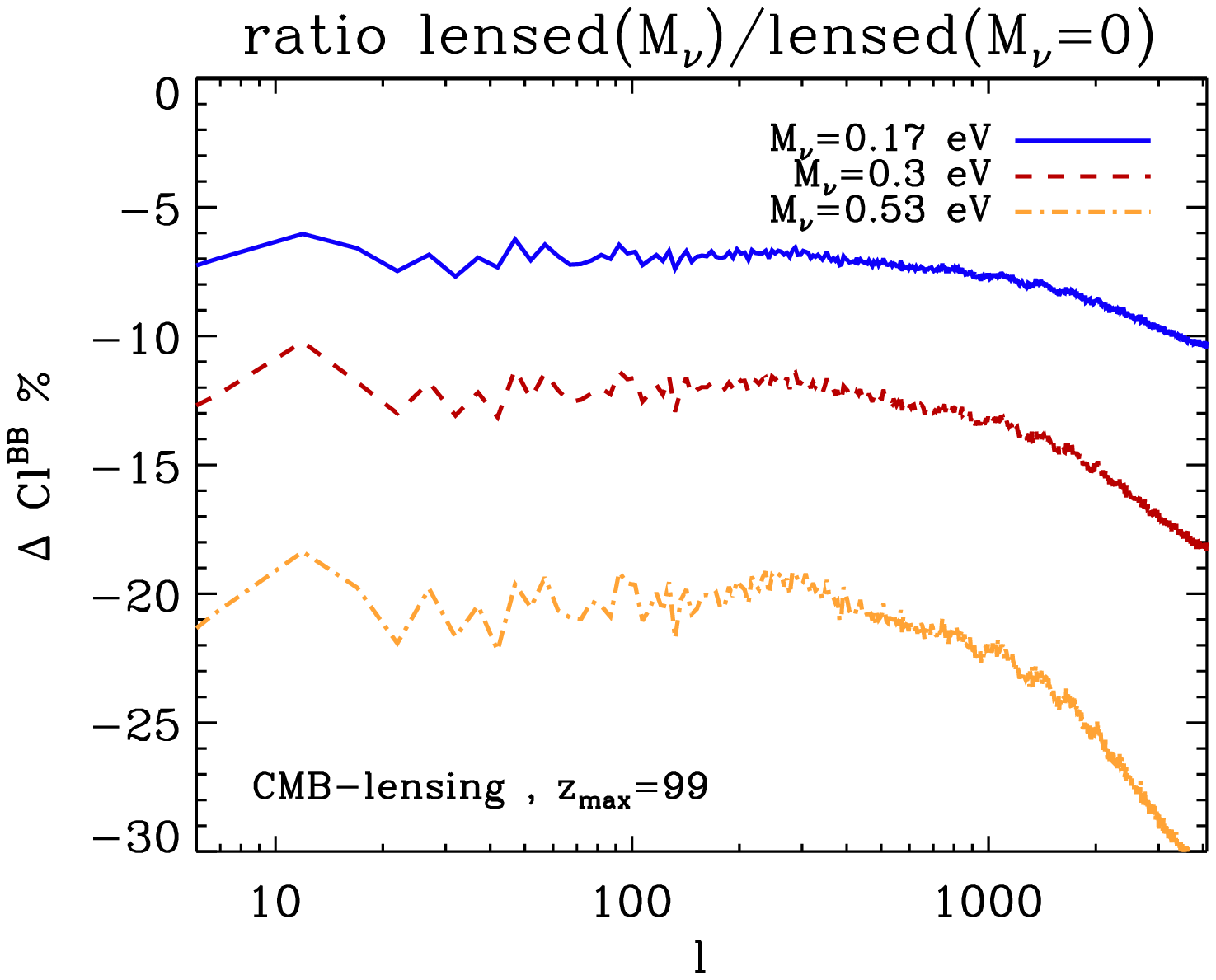}
\end{tabular}
\end{center} 
\caption{Upper left panel: CMB lensing potential angular
power spectrum obtained by photon ray-tracing, from $z=0$ up to
$z=99$, across the total gravitational field of the DEMNUni
simulations. The dashed light-green curve represents the prediction
from \CAMB\ in the massless case; the dotted black, blue, red, and
orange lines represent the simulated lensing potential for
$M_\nu=0,0.17,0.3,0.53$ eV, respectively. 
Top right panel: percent residuals
between the massive and massless neutrino cases. Here, tri-dot-dashed blue, red, and
orange lines are the simulated signals for $M_\nu=0,0.17,0.3,0.53$ eV, respectively.
Semi-analytical non-linear \CAMB\ predictions are represented by symbols, as
described in the legend. 
Middle left panel: percent residuals, wrt to the massless case, of the simulated TT lensed power spectra for
$M_\nu= 0.17,0.3,0.53$ eV. 
Middle right panel: percent residuals, wrt to the massless case, of the simulated EE lensed power spectra for
$M_\nu= 0.17,0.3,0.53$ eV.
Lower left panel: simulated lens-induced BB angular power
spectra for $M_\nu=0,0.17,0.3,0.53$ eV.
Lower right panel: percent residuals, wrt to the massless case, of the
simulated lens-induced BB power spectra for
$M_\nu= 0.17,0.3,0.53$ eV.}
\label{fig_cmblens}
\end{figure*}

\subsection{CMB and Weak lensing spectra}
Let us now focus the discussion on the effect of massive neutrinos on
lens-induced CMB secondary anisotropies. Since lensing traces
directly the total matter power spectrum, \ie the
gravitational potential $\Phi$ rather than the time derivative
$\dot{\Phi}$, the theoretical linear predictions from
\CAMB, combined with the \Halofit\ \cite{Smith_etal_2003} non-linear
corrections and including also the neutrino contribution from
\cite{Bird_etal2011}, succeed in reproducing mostly perfectly (as
compared to the simulated signal) the CMB-lensing
effects in the presence of massive neutrinos on all the scales of interest.
This result is presented in the top panels of Fig.~\ref{fig_cmblens};
in particular, the left panel shows the CMB lensing potential angular
power spectrum obtained by photon ray-tracing, from $z=0$ up to
$z=99$, across the total gravitational 
field, generated both by CDM and massive neutrinos, from the DEMNUni
simulations. The dashed light-green curve represents the prediction
from \CAMB\ in the $\Lambda$CDM case; the dotted black, blue, red, and
orange lines represent the simulated lensing potential for
$M_\nu=0,0.17,0.3,0.53$ eV, respectively. Comparing the black and
light-green lines, \ie the $\Lambda$CDM curves, we observe
less power in the simulated spectrum with respect to
\CAMB\ predictions, at multipoles $l\gtrsim 1000$. This is due to the finite resolution of the
DEMNUni potential grids, about $\sim 0.5$ Mpc/$h$, much smaller
compared to the resolution of the dark matter simulations from
\cite{Takahashi_etal_2012} used for non-linear corrections in
\Halofit. On the other hand, looking at the top
right panel of Fig.~\ref{fig_cmblens}, \ie at the percent residuals
between the massive and massless neutrino cases, we do not observe
such lack of power with respect to the semi-analytical
\CAMB\ predictions (compare tri-dot-dashed lines against symbols), 
implying that resolution effects cancel out
when focusing on the relative differences with respect to the
$\Lambda$CDM case. In particular, as expected, the integrated effect of
massive neutrinos produces a suppression of power in the
lensing potential
which increases with the neutrino mass, decreases with the
angular scale (low $l$), and at large multipoles tends to become
constant, approximately proportional to the ratios of the total
neutrino masses. We find an asymptotic power suppression of about $\Delta
C_l^{\phi\phi}\simeq 10\%, 19\%, 31\%$ for $M_\nu=0.17,\,0.30,\,0.53$ eV,
respectively. As mentioned above, this trend is directly related to 
the behaviour of the total matter power spectra shown in
Fig.~\ref{pk}, as the lensing potential power spectrum can be written as the integral along the line
of sight of the matter power spectrum weighted by a geometrical
factor.

In the middle and lower panels of Fig.~\ref{fig_cmblens} we show the
effect of massive neutrinos on the lensed CMB temperature (TT), lensed E-mode
polarisation (EE), and lens-induced B-mode polarisation (BB) angular
power spectra. As in
Refs.~\cite{Carbone_etal_2008,Calabrese_etal_2015},
the lensed power
spectra have been obtained by modifying the
\LensPix\ code\footnote{http://cosmologist.info/lenspix/}\cite{Lewis2005},
in order to use directly as input the power spectra and phases of the
lensing potential
maps produced by ray-tracing across the DEMNUni
simulations. As expected, the percent residuals between the
massive and massless neutrino cases, presented in the middle panel of
Fig.~\ref{fig_cmblens}, show that free-streaming neutrinos alter the lensed TT
and EE power spectra. In particular, as the strength of the
gravitational potential decreases for larger neutrino masses, CMB
acoustic oscillations are less lensed, {\emph i.e.} they
are less smoothed and smeared out than in the massless case. This implies
that the larger the neutrino mass is, the larger the lensed TT and EE
power spectra are at large angular scales (small $l$), and the smaller
the lensing amplification of the so-called ``damping-tail'' at
small scales (high $l$) is. As a consequence, the production of the lens-induced B-mode
polarisation, via rotation of the E-mode pattern, is less enhanced in
the presence of massive neutrinos, and the amplitude of the BB power
spectrum decreases with increasing neutrino masses on all the
scales, as shown in the left lower panel of Fig.~\ref{fig_cmblens}. 
As the right panel shows, the B-mode lack of
power due to the presence of massive neutrinos is mostly scale
independent (except at $l>1000$ where  non-linear effects come
into play) \cite{Oyama2013, Oyama2016}. The reason is well known, {\emph i.e.} B-modes are due to
the transfer of E-mode power from small scales to large scales, and on
scales $1000<l<2000$ the effect of massive neutrinos on E-modes is
almost scale-independent, as the right middle panel of
Fig.~\ref{fig_cmblens} shows. Again, the B-mode power residuals
are somewhat proportional, for $l\rightarrow 0$, to the ratio of the neutrino masses
assumed in the simulations.

\begin{figure*}[!ht]
\begin{center}
\setlength{\tabcolsep}{0.01pt}
\begin{tabular}{c c} 
\includegraphics[width=0.5\textwidth]{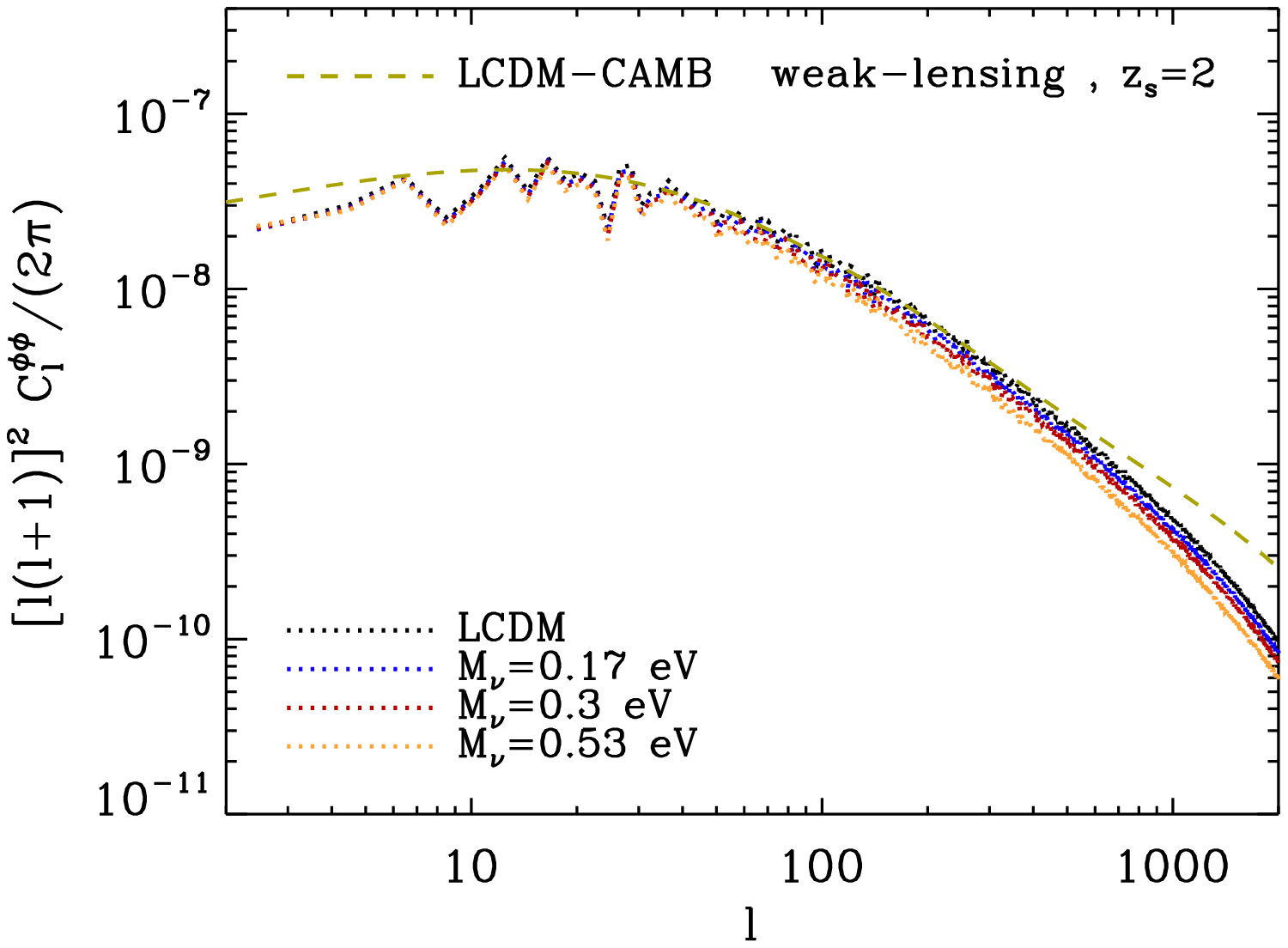}&
\includegraphics[width=0.5\textwidth]{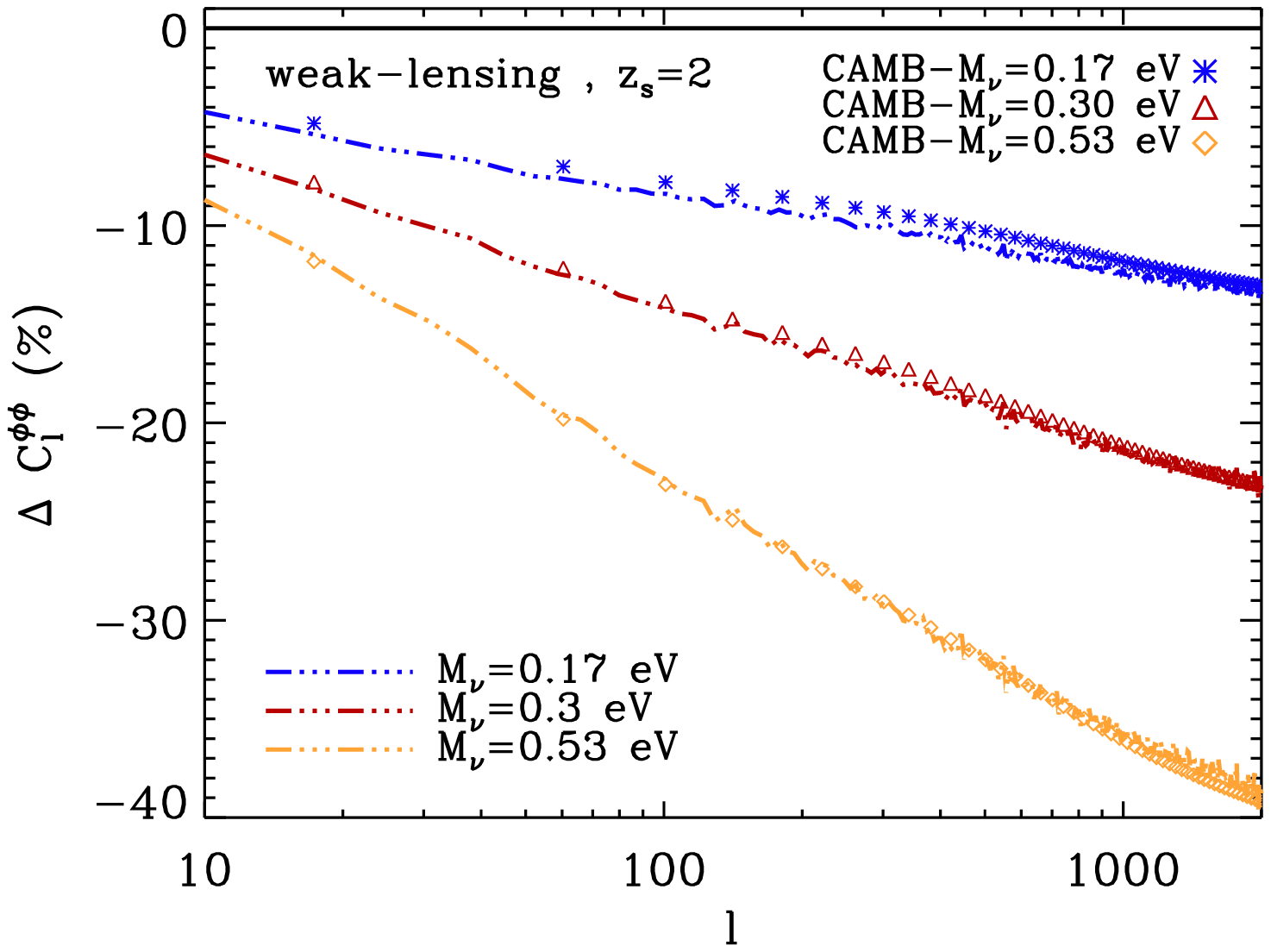}\\
\includegraphics[width=0.5\textwidth]{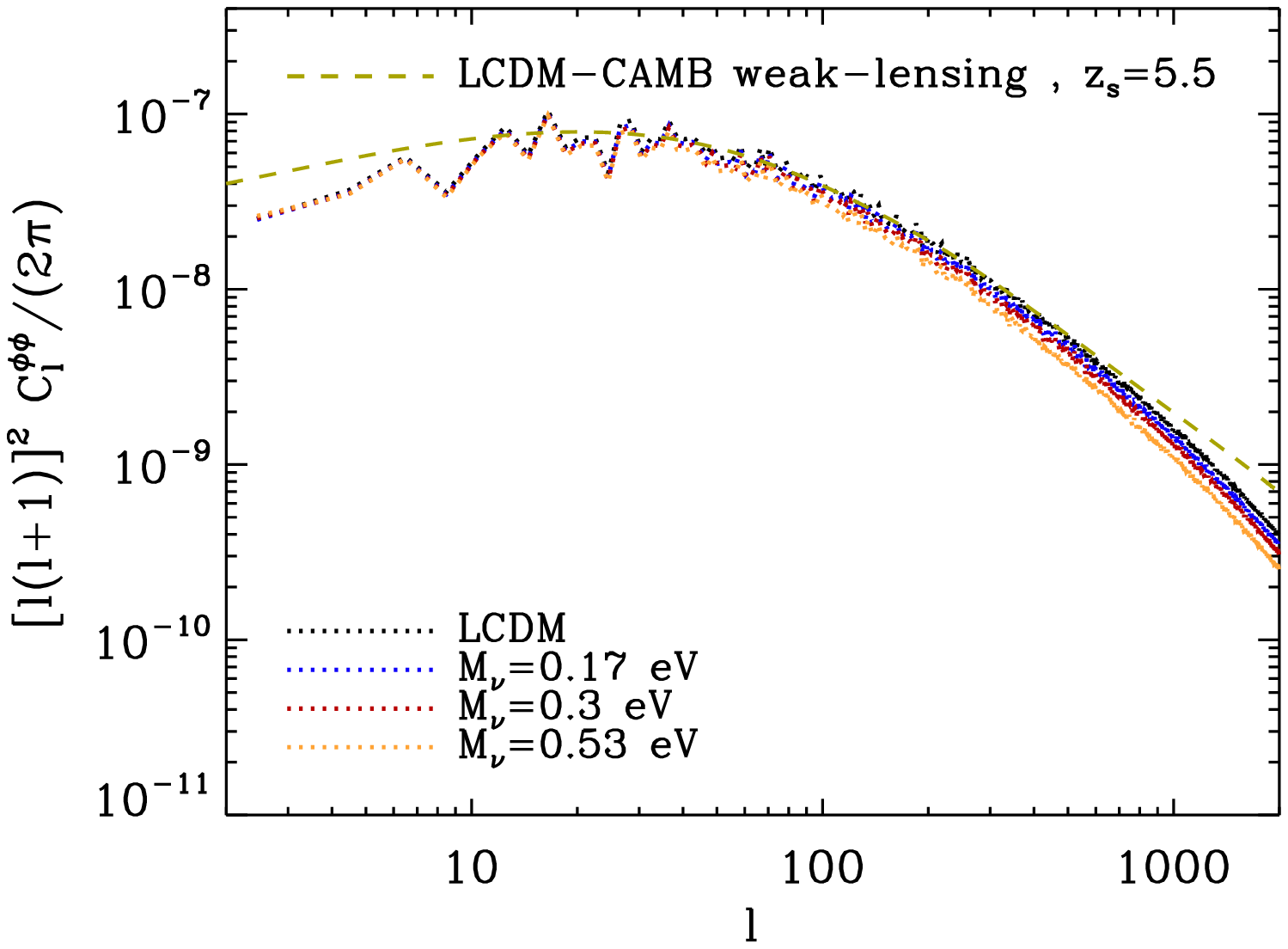}&
\includegraphics[width=0.5\textwidth]{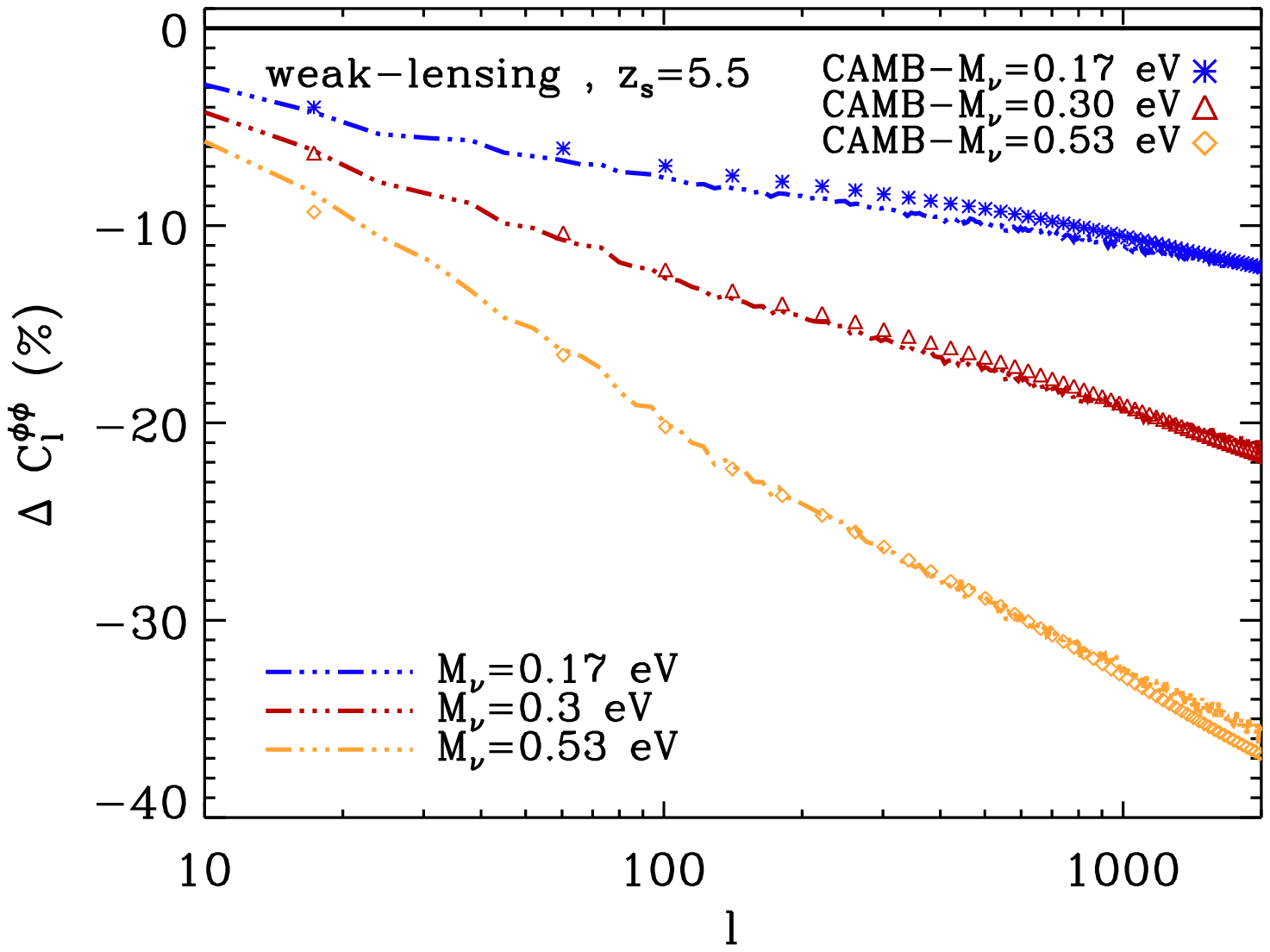}\\
\includegraphics[width=0.5\textwidth]{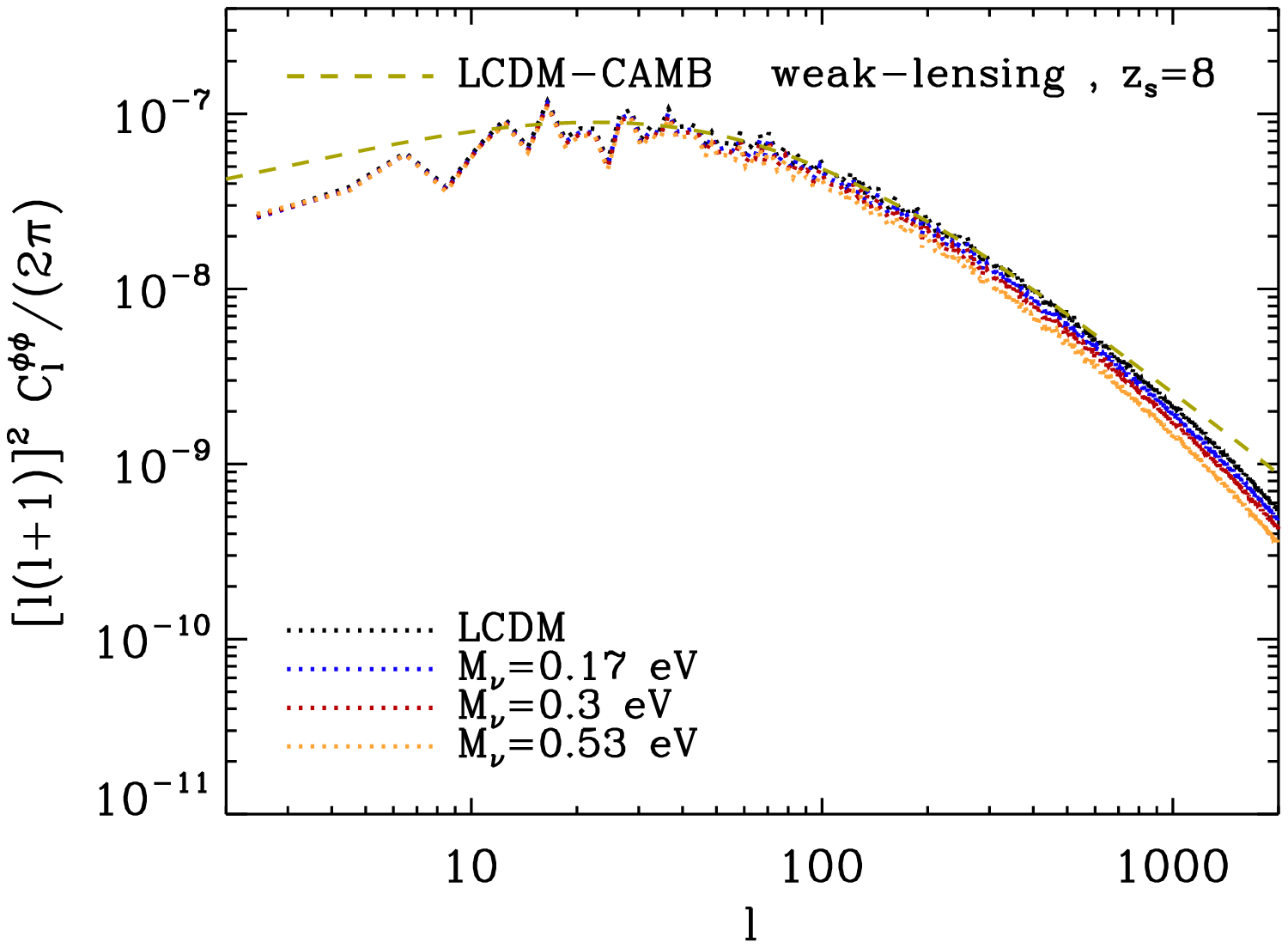}&
\includegraphics[width=0.5\textwidth]{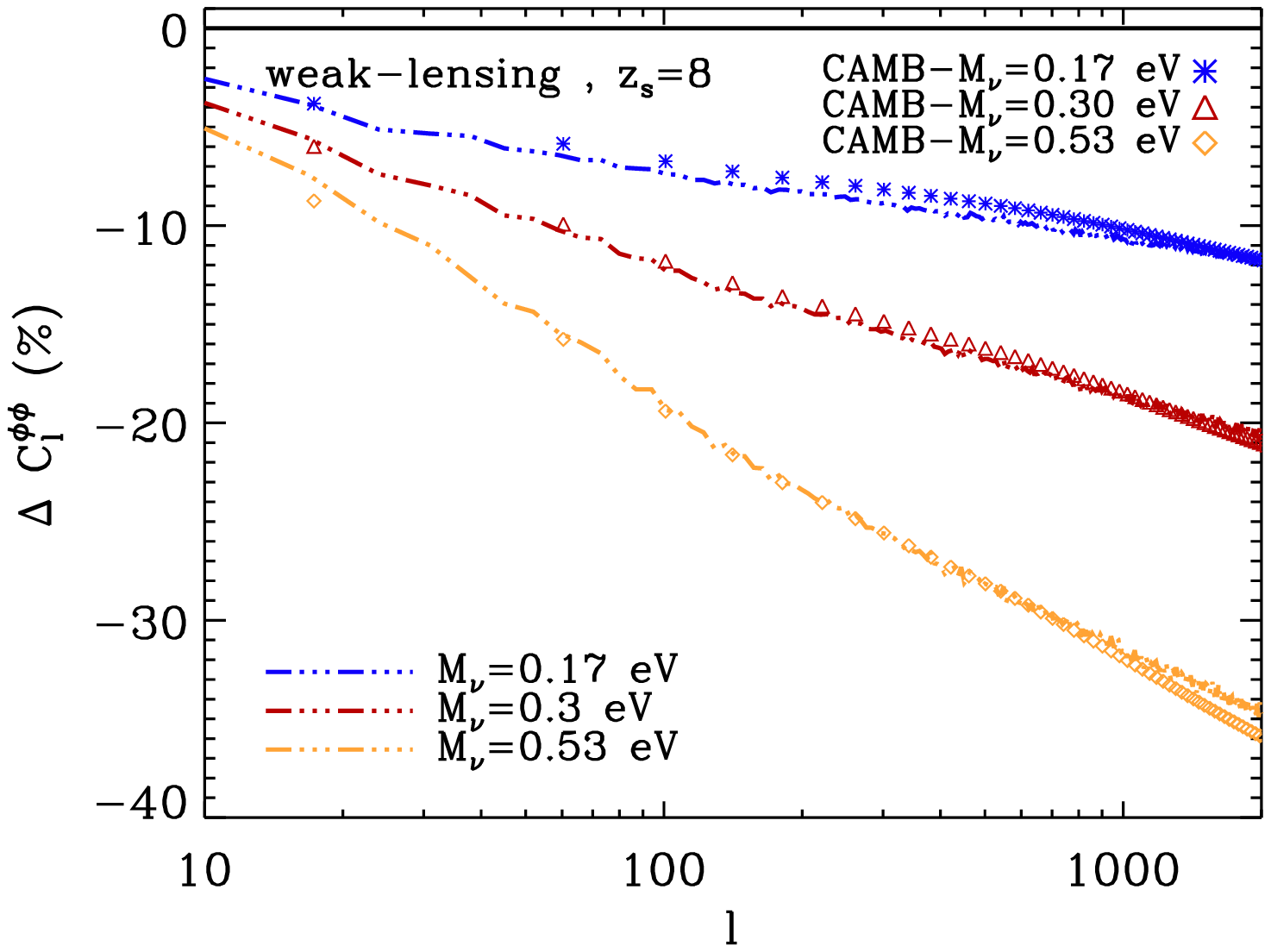}
\end{tabular}
\end{center} 
\caption{Left: lensing potential angular power spectra for
  sources all placed at redshifts $z=2$ (upper panel),
$z=5.5$ (middle panel), and $z=8$ (lower panel), respectively. The dotted black, blue, red and
  orange lines represent the simulated signals for
  $M_\nu=0,0.17,0.3,0.53$ eV, respectively. 
The $\Lambda$CDM non-linear expectations from \CAMB\ are represented
by the dashed green line.
Right: corresponding percent residuals wrt the massless
case. Here, tri-dot-dashed blue, red, and
orange lines are the simulated signals for $M_\nu=0,0.17,0.3,0.53$ eV, respectively.
Semi-analytical non-linear \CAMB\ predictions are represented by symbols, as
described in the legend.}
\label{fig_weaklens}
\end{figure*}

Finally, in Fig.~\ref{fig_weaklens} we present the results for three
cases of weak-lensing, with sources all placed at redshifts $z=2$,
$z=5.5$, and $z=8$, respectively. As expected, in all the cases, we find the
lensing potential to behave in a way very similar to the CMB lensing potential. In
fact, its angular power spectrum is suppressed by
free-streaming massive neutrinos, and such suppression is scale
dependent, increasing with the multipoles $l$, as new scales $k$
exceed the free-streaming scale $k_{fs}$. We have modified \CAMB\ to
account for the computation of the weak-lensing signal for sources
placed all at the same redshift $z_s$, by substituting $z_{\rm LS}$ with
$z_s$ in the \CAMB\ routine ``equations.f90'' for the calculation of
the ``sources(3)'' variable; the corresponding analytical
expectations are represented by the dashed green line in the left
panels of Fig.~\ref{fig_weaklens}. As expected, given the finite
simulation volume, on large scales the simulated signal recovers
much better CAMB predictions as redshift decreases, while on small
scales we observe an increase of the lack of power due the limited
resolution, $\sim 0.5$ Mpc/$h$, of the gravitational potential grids used for
ray-tracing. Nonetheless, when looking at the relative differences
between the massive and massless cases, for different $M_\nu$ values, 
the simulated signal recovers the expectations within
the $\sim 1$\% accuracy level (see the right panel of Fig.~\ref{fig_weaklens}). 
This means that the effect of massive neutrinos on LSS formation
decouples from highly non-linear regime physics, as {\emph e.g.}
baryon effects, since, as well known, massive neutrinos escape the
potential field of small scale structures, causing such effects 
to cancel out when considering relative differences. 

\section{The impact of massive neutrinos on cross power spectra}
\label{cross-power}
Let us now consider the effects of massive neutrinos on
the cross angular power spectra between CMB/weak-lensing and the
ISW-RS signals. Since the ISW-RS effect is not directly observable, 
the ISW-RS cross-correlation with weak-lensing 
(together with its cross-correlation with galaxies, which we do not
discuss in this work) 
allows to observe and measure the impact of
time-varying potentials on light travelling to us from the last
scattering surface.
It is worth noting that here, for the first time in the literature, we
present the non-linear behaviour of such cross-correlation signal, as extracted from
N-body simulations accounting for free-streaming massive
neutrinos. The linear counterpart can be computed using Boltzmann
codes as \CAMB\ or \CLASS\footnote{http://class-code.net/} \cite{CLASS}.

\subsection{ISW-RS--CMB-lensing cross-correlations}
\label{ISWRS--CMB-lensing cross}
\begin{figure*}[!ht]
\begin{center}
\setlength{\tabcolsep}{0.01pt}
\begin{tabular}{c c}
\includegraphics[width=0.5\textwidth]{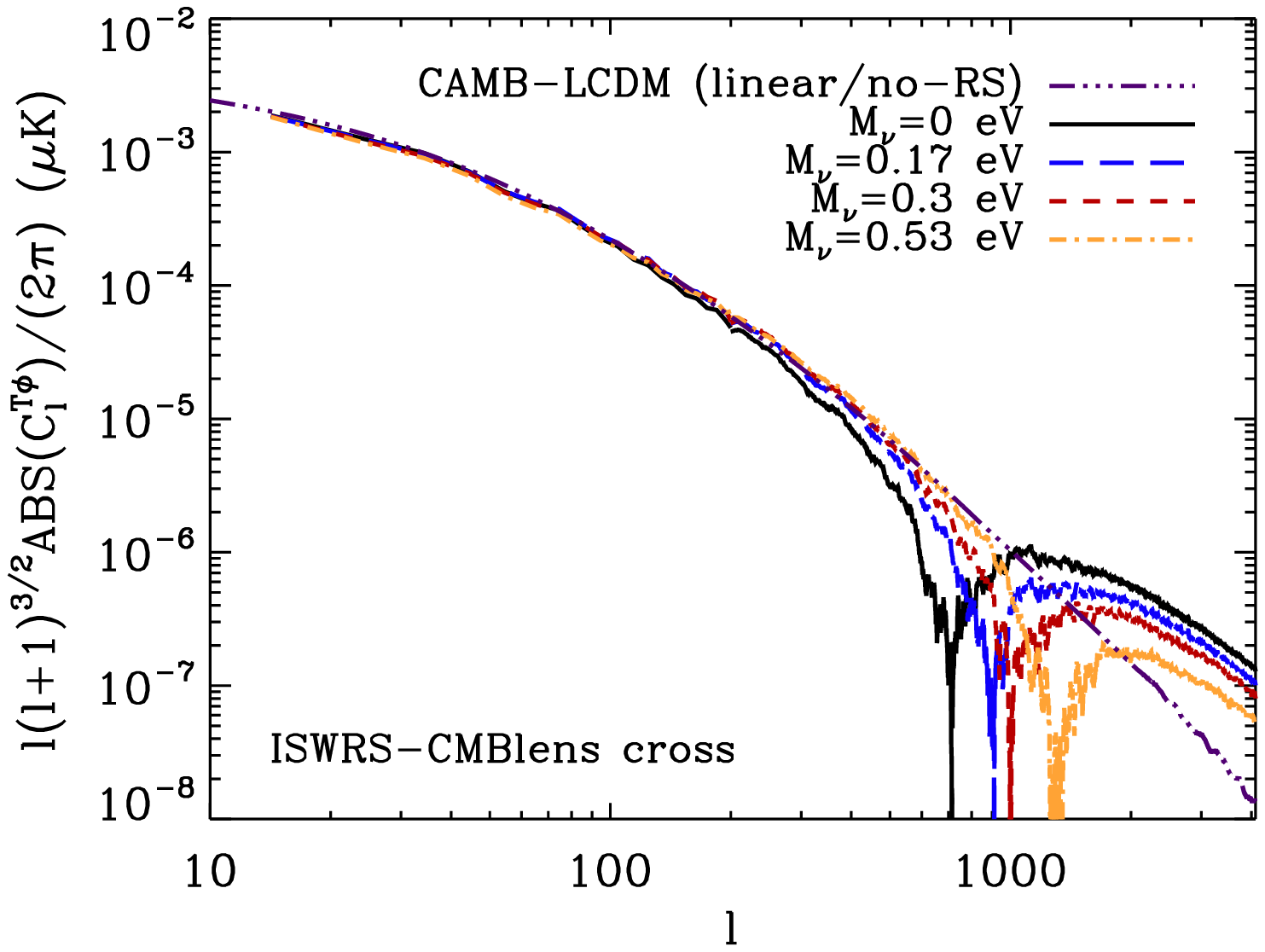}&
\includegraphics[width=0.5\textwidth]{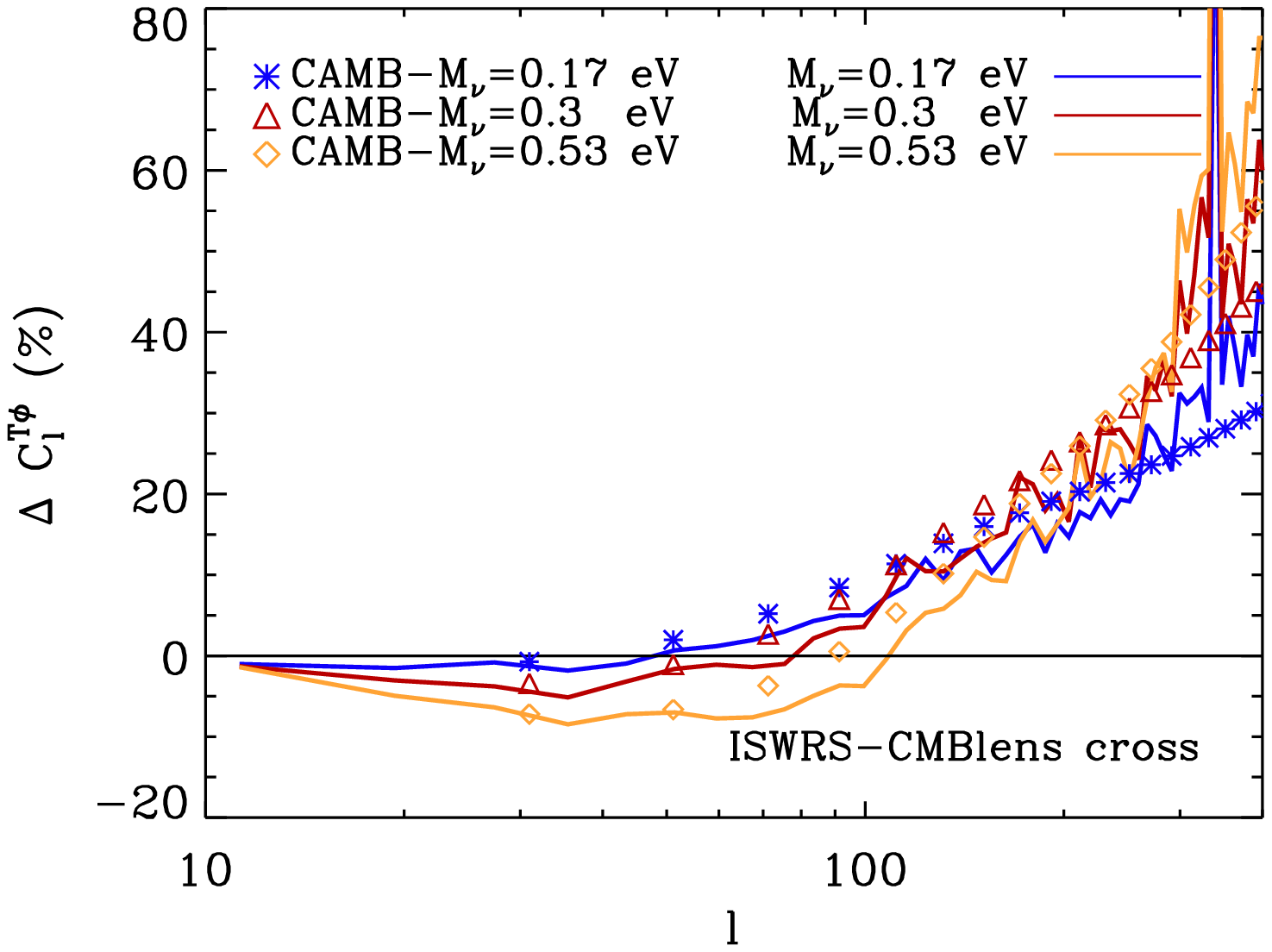}\\
\includegraphics[width=0.5\textwidth]{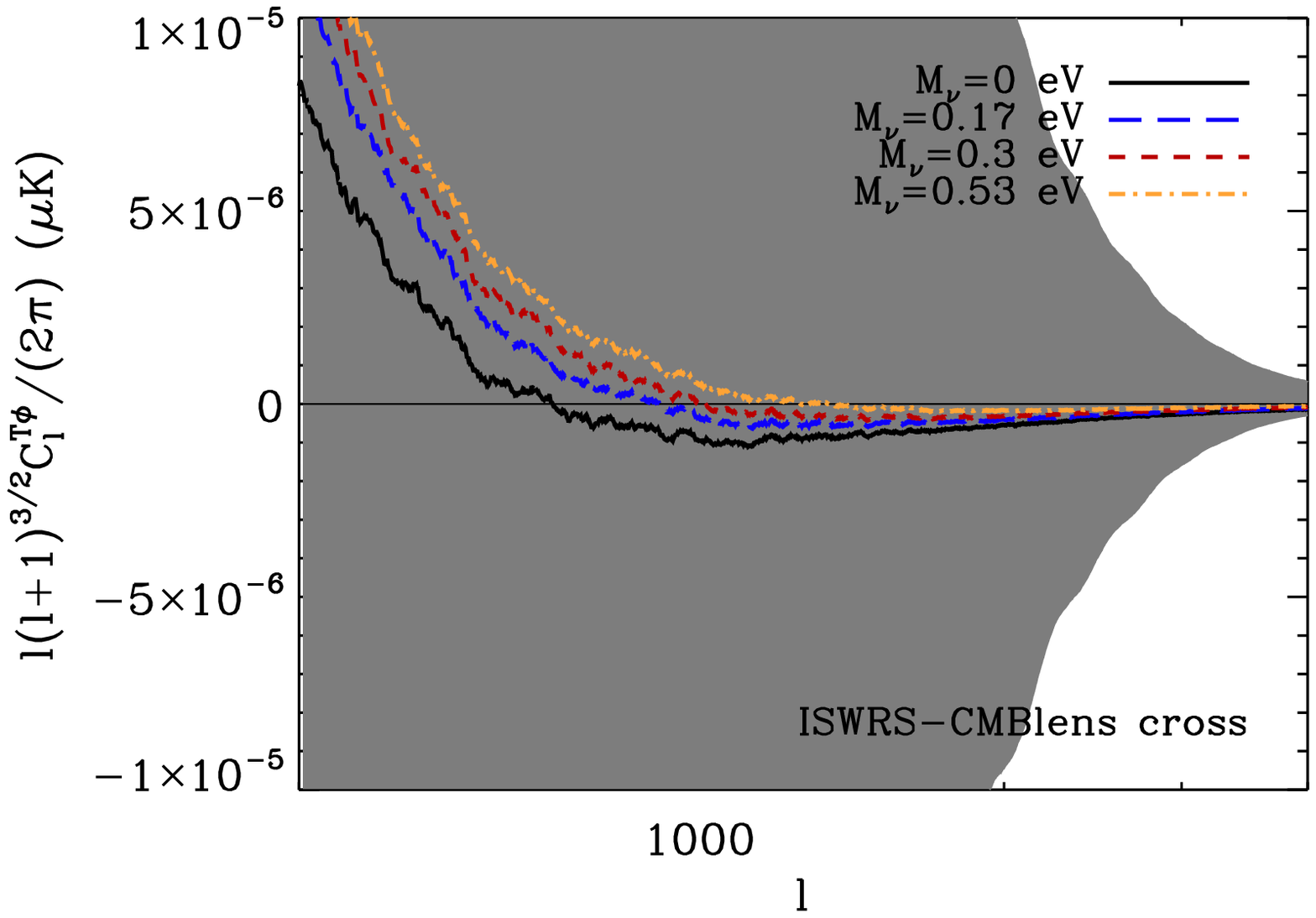}&
\includegraphics[width=0.5\textwidth]{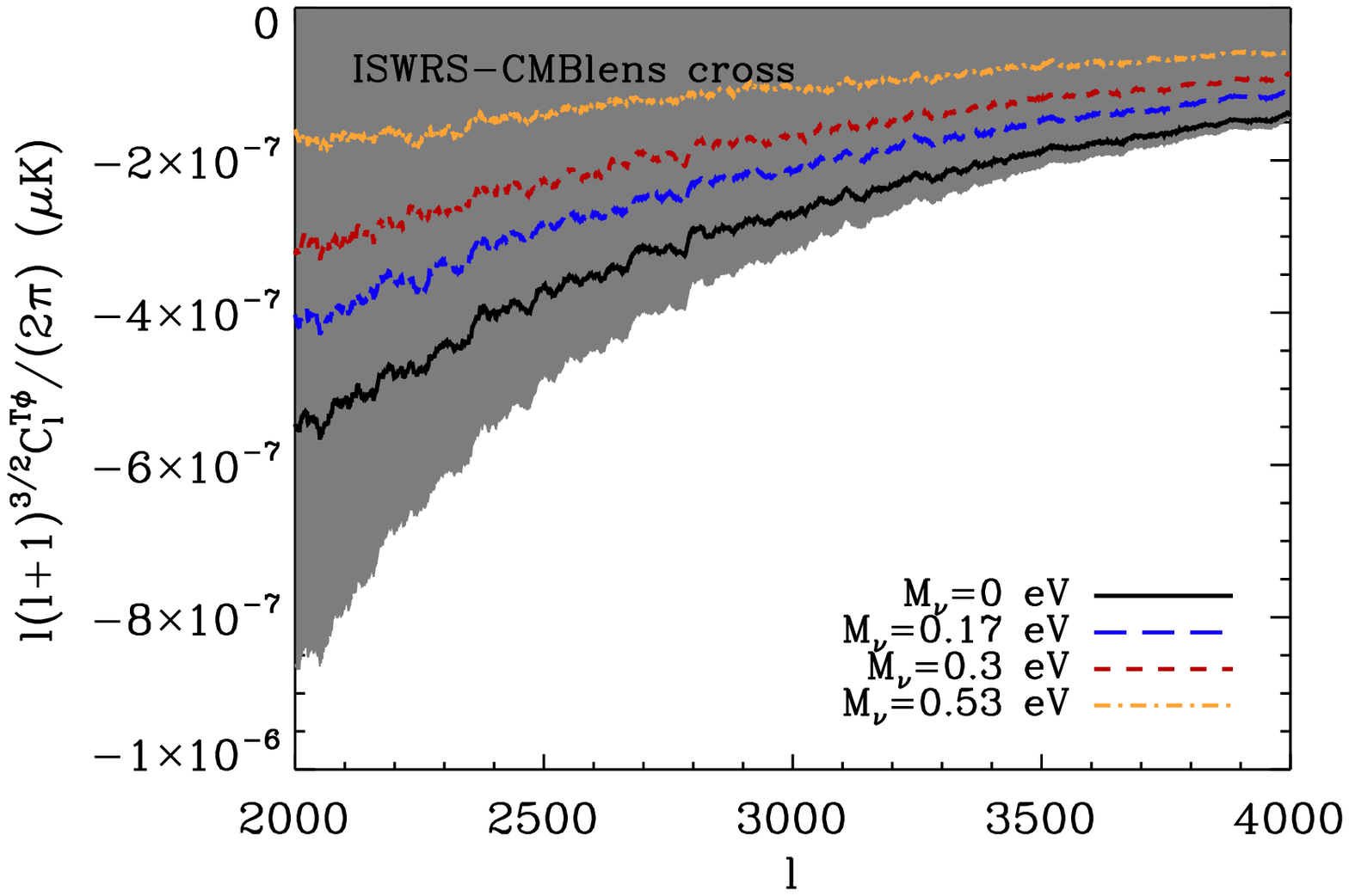}
\end{tabular}
\end{center} 
\caption{Left upper panel: absolute values of the simulated cross power spectra between 
CMB-lensing and  the total ISW-RS induced temperature
anisotropies, obtained via ray-tracing from $z=0$ to $z\simeq 21$,
for $M_\nu=0,0.17,0.3,0.53$ eV (solid black, long-dashed blue, dashed
red, and dot-dashed orange lines, respectively).  The violet
tri-dot-dashed line represents the linear contribution from \CAMB\ in the
massless case.
Right upper panel: corresponding percent residuals wrt the massless
case on linear scales $l<400$. Here, blue, red, and
orange lines are the simulated signals for $M_\nu=0,0.17,0.3,0.53$ eV, respectively.
Semi-analytical linear \CAMB\ predictions are represented by symbols, as
described in the legend.
Bottom panels: simulated cross power
spectra for $l>400$.  The shaded
grey area represents the cosmic variance associated to the signal.}
\label{fig_cmbl_isw_cross}
\end{figure*}

The absolute values of the simulated cross power spectra between 
CMB-lensing and  the total ISW-RS induced temperature
anisotropies, obtained via ray-tracing from $z=0$ to $z\simeq 21$, are
shown in the left panel of Fig.~\ref{fig_cmbl_isw_cross} for
four different total neutrino masses, $M_\nu=0,0.17,0.3,0.53$ eV. They
are represented by the black (solid), blue (long-dashed), red (dashed),
and orange (dot-dashed) lines, respectively. The violet
tri-dot-dashed line represents the linear contribution from
\CAMB\footnote{As already mentioned in \ref{ISWRS}, at present Boltzmann codes are
not able to compute the non-linear contribution to $\dot{\Phi}$.} in
the massless case. Above all, let us observe that, using a finite box of
$2$ Gpc/$h$, we manage to recover the predicted linear signal starting
from very low multiples, $l\sim 10$. This is a confirmation of the
accuracy of the technique implemented to extract the ISW effect from the
DEMNUni simulations.
In addition, at the transition from
linear to non-linear scales, we recover also the expected sign inversion of the
cross-correlation spectra due to the negative
\emph{non-linear} correlation between the RS effect and matter density, which becomes dominant with respect to the net positive \emph{linear} correlation between
density and CMB temperature, produced by the decay of the
\emph{linear} gravitational potential, as the Universe expands in the
presence of dark energy \cite{Smith_etal_2009, Schaefer2011}.  
This leads to a sign change of the cross-spectrum, which for a $\Lambda$CDM
model with the same cosmological parameters as in the
simulations, happens to be about at $l\sim 700$ \cite{Mangilli13, Lewis_RS}.

As the left top panel of Fig.~\ref{fig_cmbl_isw_cross} shows, 
the major effect of massive neutrinos on the CMB-lensing/ISW-RS cross
spectrum consists of moving the sign inversion position
toward larger multipoles, producing a larger displacement as the
neutrino mass increases.  This can be explained considering
that the larger the neutrino mass is, the larger the suppression of
structure formation is, and therefore cosmological perturbations tend to
stay in the linear regime on smaller scales than in the massless
case. This implies that free-streaming massive neutrinos not only
produce an excess of ISW-RS power, but also an excess of cross-correlation between CMB-lensing and the ISW-RS effect, and
this time such excess increases with larger neutrino masses on scales
  $100<l<1000$, being a factor of $\sim 4$ for $M_\nu=0.3$ eV, at
$l\sim 600$. On multipoles $700<l<1000$, the shift of the sign inversion
  is the  the dominant feature, and, indeed, a future detection and
  measurement of its position could be a further probe of the total
  neutrino mass. On larger multipoles the
  asympthotic suppression of the matter power spectra due to neutrino
  free-streaming becomes dominant, and we recover
the usual trend,  {\emph{i.e.}} larger neutrino masses  produce a larger decrease of
  the non-linear cross power spectrum.

In the right top panel of Fig.~\ref{fig_cmbl_isw_cross} we compare
our findings with \CAMB\ predictions on linear scales.  On
multipoles $10<l<400$ the accuracy of the reconstructed signal is
very high, about $1-2$\%, and the residuals start to
increase only when non-linear effects come into play, {\emph{i.e}} at
$l>400$. On such scales, the behaviour of the simulated cross power
spectra is shown in the bottom panels of
Fig.~\ref{fig_cmbl_isw_cross}. In particular, the different curves in left panel
represent the sign inversion due to non-linearities, and the shaded
grey area the associated cosmic variance $\sqrt{[(C_l^{T\kappa})^2+C_l^{TT} C^{\kappa\kappa}_l]/(2l+1)}$, which unfortunately makes
measurements of this effect quite challenging, since the primary CMB temperature
anisotropies act as a foreground in this case.

\begin{figure*}[!ht]
\begin{center}
\setlength{\tabcolsep}{0.01pt}
\begin{tabular}{c c} 
\includegraphics[width=0.5\textwidth]{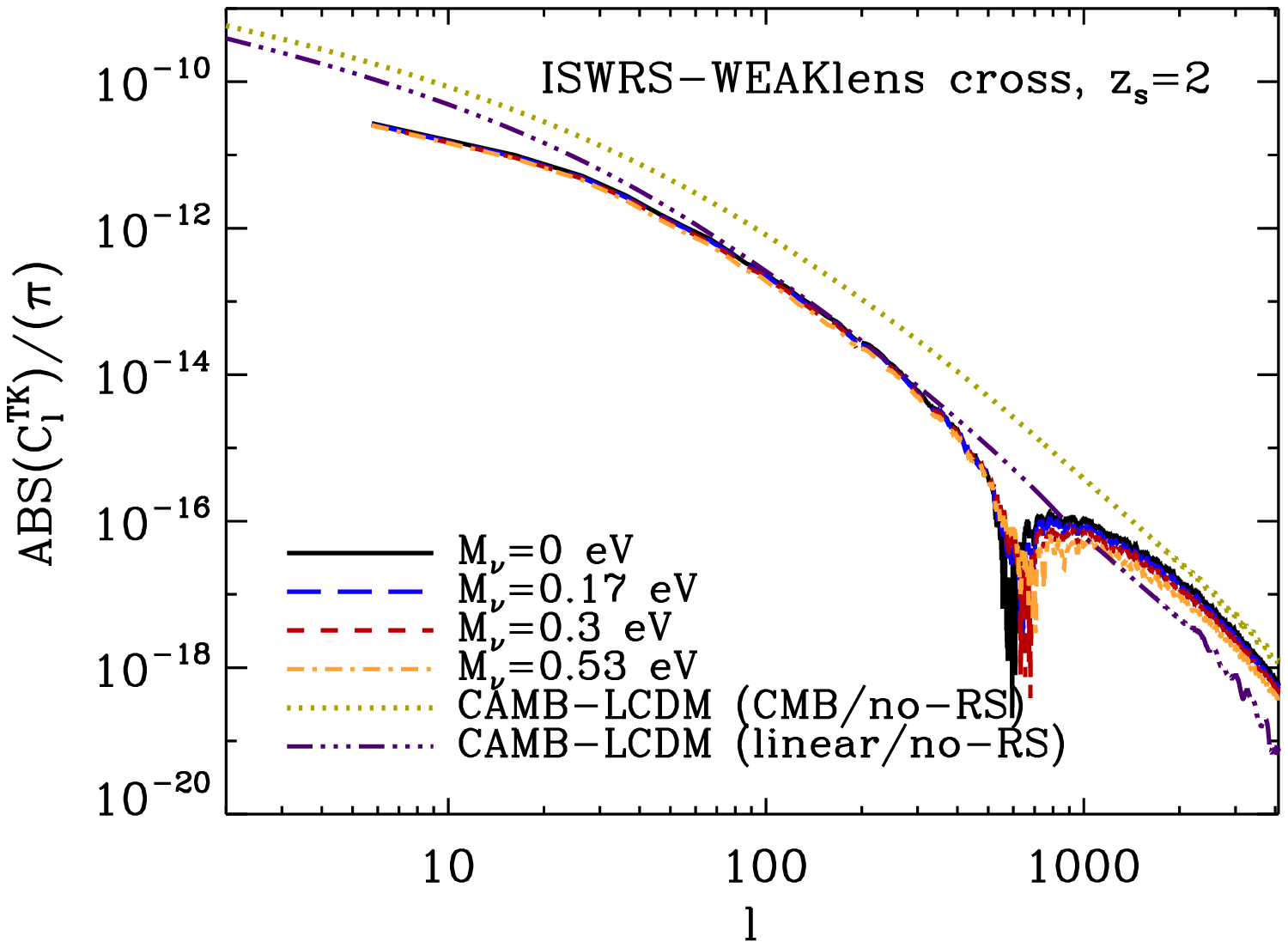}&
\includegraphics[width=0.5\textwidth]{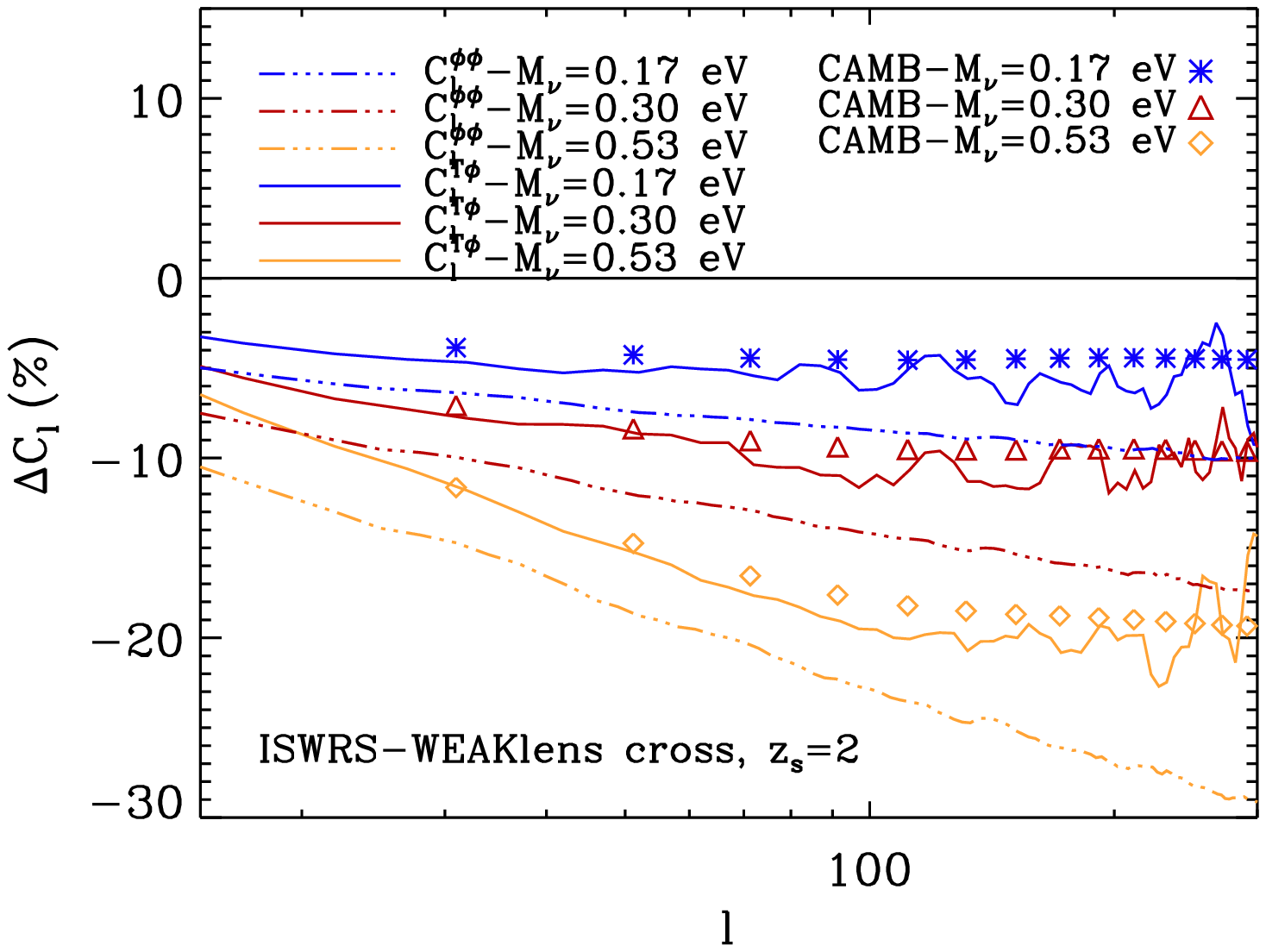}\\
\includegraphics[width=0.5\textwidth]{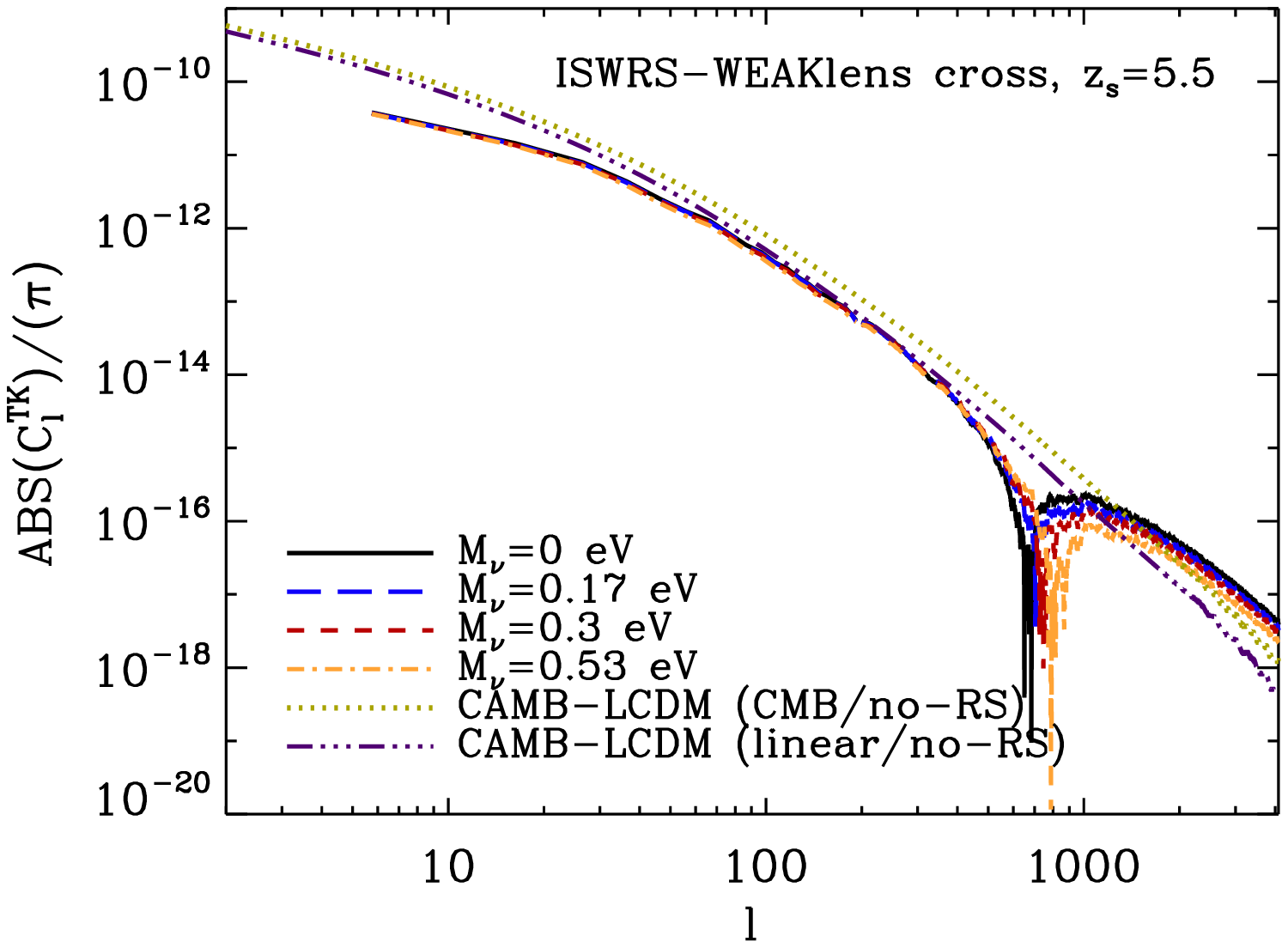}&
\includegraphics[width=0.5\textwidth]{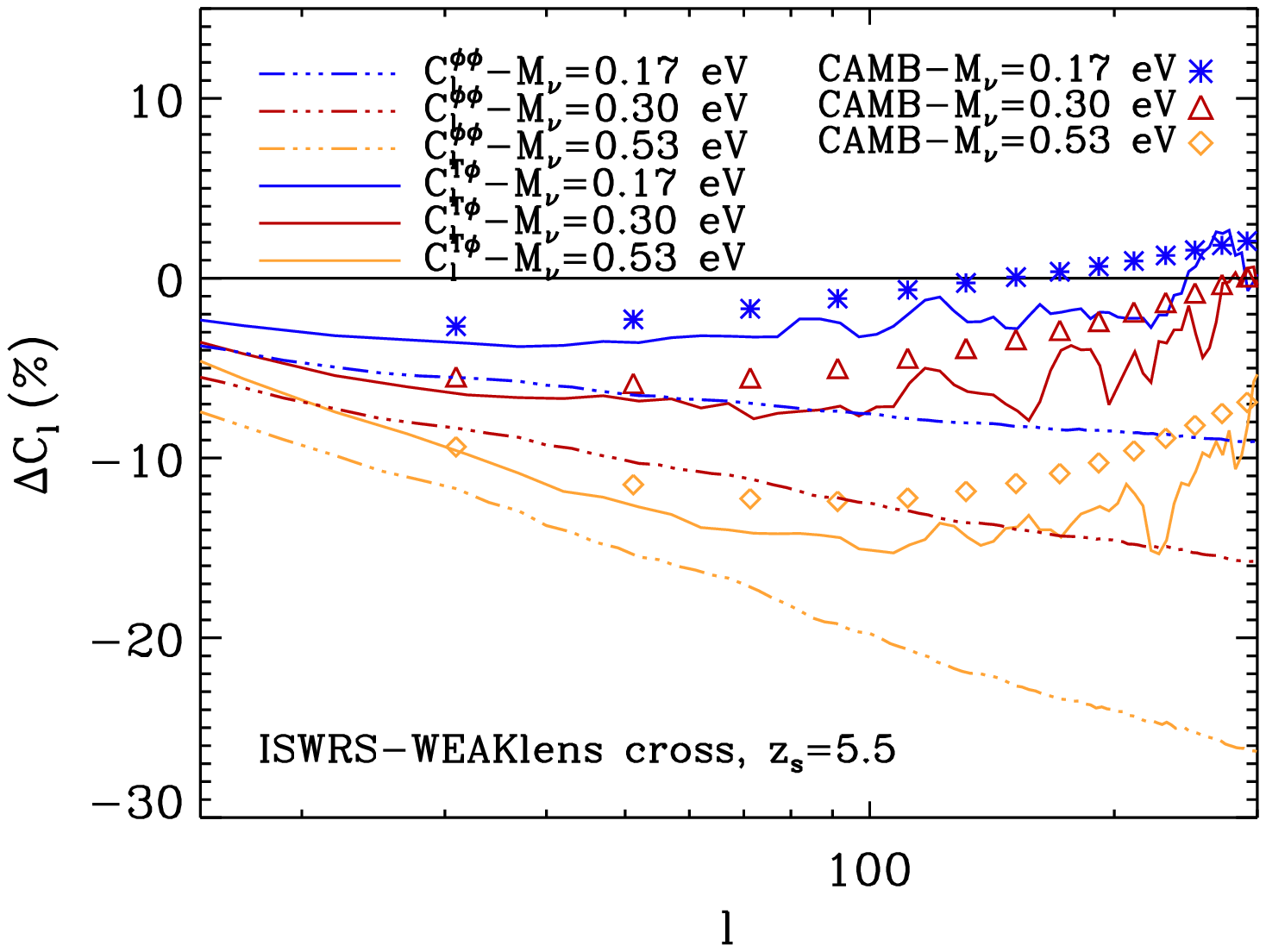}\\
\includegraphics[width=0.5\textwidth]{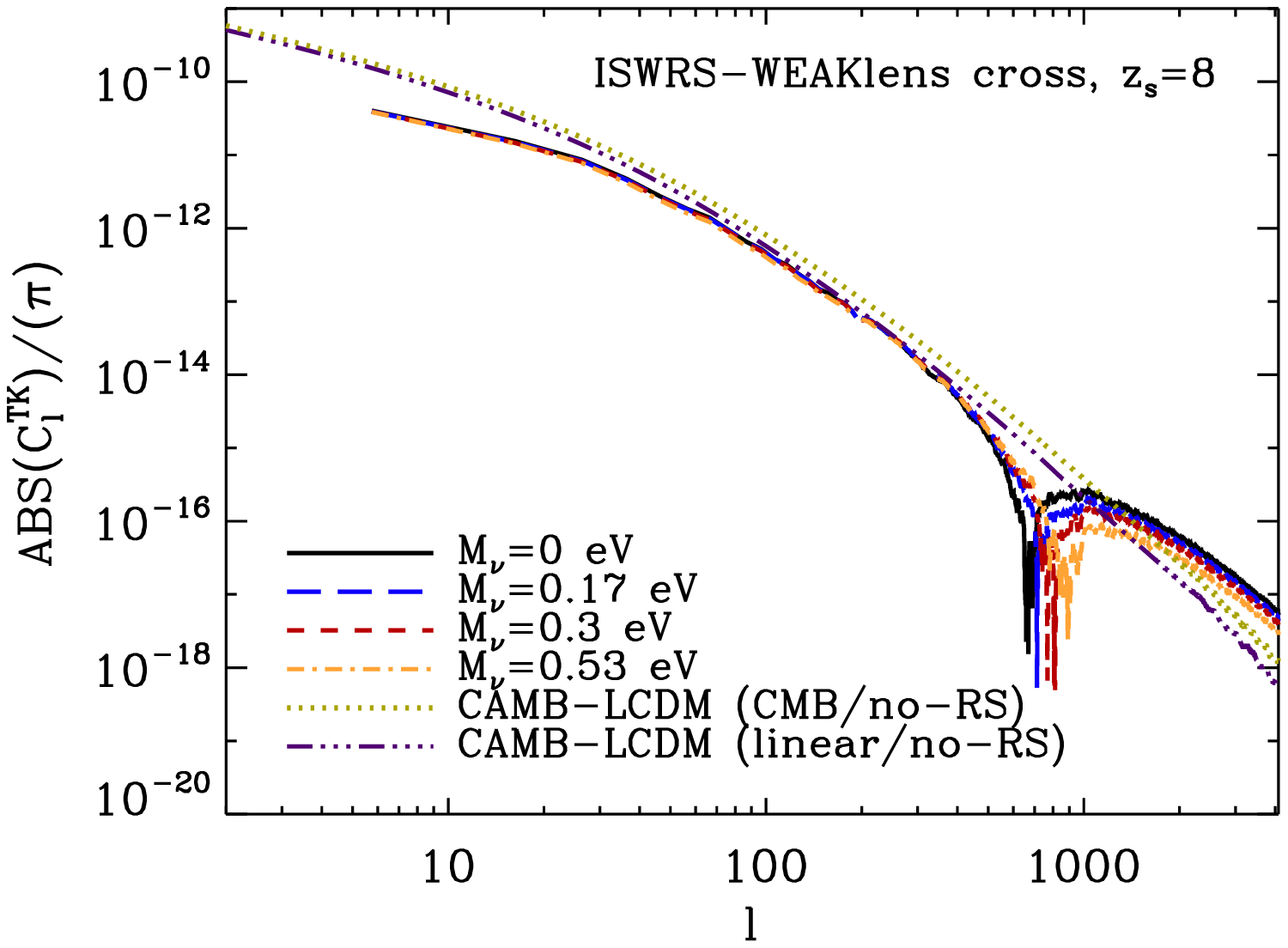}&
\includegraphics[width=0.5\textwidth]{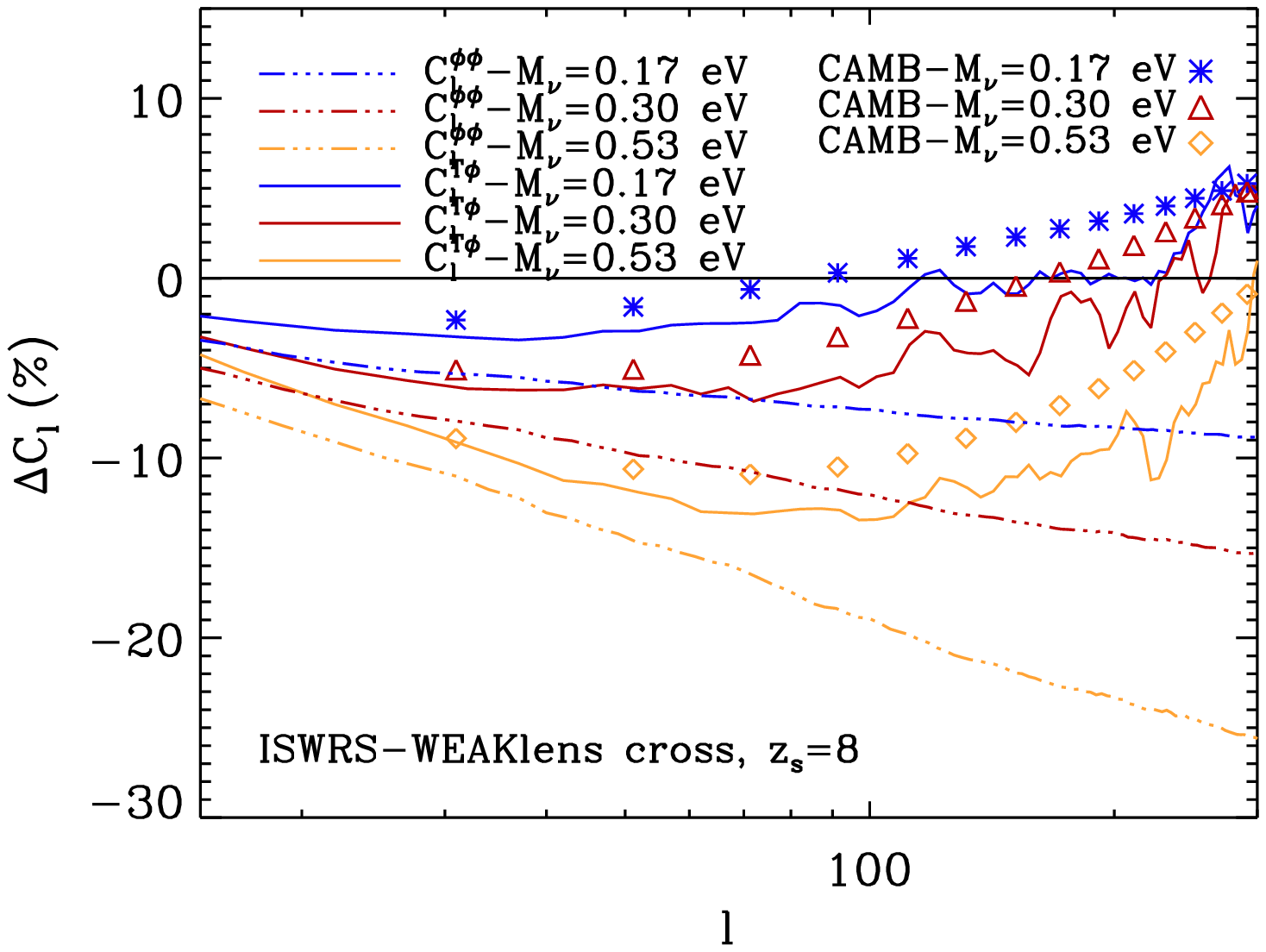}
\end{tabular}
\end{center} 
\caption{Left: absolute values of the WL/ISW-RS angular
  cross spectra for  sources placed at redshifts $z=2$ (upper panel),
$z=5.5$ (middle panel), and $z=8$ (lower panel), respectively. The
  solid black, long-dashed blue, dashed red, and dot-dashed
  orange lines represent the simulated signals for
  $M_\nu=0,0.17,0.3,0.53$ eV, respectively. The $\Lambda$CDM linear expectations from \CAMB\ are represented
by the tri-dotted violet line. The light-green dotted line is the linear
ISW-RS/CMB-lensing cross spectrum from \CAMB, here shown for
comparison. 
Right: percent residuals wrt the massless
case for $l<300$. Tri-dot-dashed blue, red, and
orange lines correspond to the simulated signals of the auto WL power
spectra for $M_\nu=0,0.17,0.3,0.53$ eV, respectively.
Solid blue, red, and orange lines are the residulars of the simulated
ISW-RS/WL cross spectra. The corresponding
semi-analytical linear \CAMB\ predictions are represented by symbols, as
described in the legend.}
\label{fig_weakl_isw_cross}
\end{figure*}

\subsection{ISW-RS--weak-lensing cross-correlations}
\label{ISWRS--WEAK-lensing cross}

Finally, we consider the cross-correlation between the ISW-RS and the
weak-lensing signals.  The left panels of
Fig.~\ref{fig_weakl_isw_cross} show the cross spectra obtained with
lensing sources all placed at $z_s=2,\,5.5,\,8$ and for
$M_\nu=0,0.17,0.3,0.53$ eV, represented by the black (solid), blue (long-dashed), red (dashed),
and orange (dot-dashed) lines, respectively. The violet
tri-dot-dashed line represents the corresponding linear contribution from
\CAMB, while the light-green dotted line is the linear
ISW-RS/CMB-lensing cross spectrum from \CAMB, here shown for comparison. As in \S
\ref{ISWRS--CMB-lensing cross}, also in this case we find a mostly
perfect agreement between CAMB linear predictions and the simulated
signals, at scales $30 \lesssim l \lesssim 400$. On smaller multipoles,
$l\lesssim 30$ window effects take place producing a lack of power in
the simulated cross signal. As for the ISW-RS/CMB-lensing cross
spectrum, also in this case on larger multipoles, $l\gtrsim 400$,
non-linear effects produce a sign inversion, whose position stays
however mostly constant with increasing neutrino masses, even if we
notice that it is more shifted toward larger multipoles as $z_s$
increases (as expected from the theory of linear perturbation evolution).

In the right panels of Fig.~\ref{fig_weakl_isw_cross} we show the
residuals, with respect to the massless case, of the simulated
ISW-RS/weak-lensing cross spectra (solid lines), together with
linear predictions from \CAMB\ (symbols). For comparison, we also show
corresponding residuals for the simulated weak-lensing auto
spectra (tri-dot-dashed lines).  
For $z_s=2$ we find an excellent agreement within $1$\% accuracy; at
larger $z_s$ the agreement is still good with an accuracy of about
$2-3$\%. This lower accuracy is probably due to percent differences
in the ISW-RS reconstruction between \CAMB\ and our ray-tracing
technique. Overall, at $30\lesssim l\lesssim 300$ we observe a scale
dependent suppression of the cross signal which increases with the
neutrino mass. At $l\gtrsim 300$, the trend starts to reverse, and for
lighter neutrino masses the cross spectra exceed their value in the
massless case.

It is worth to note that the suppression, due to massive neutrinos, of
the ISW-RS/weak-lensing cross spectra is smaller than for the
weak-lensing auto spectra (compare solid lines against tri-dot-dashed
lines). Moreover, while in the latter case this suppression is mostly
constant with $z_s$, in the former case it decreases with the
increase of the source redshift, and finally, for very high values
of $z_s$, neutrino free-streaming produces an excess of signal with
respect to the massless case, as we observe also in the CMB case to a
greater extent. 
This is due to the larger excess of ISW-RS in the presence of massive neutrinos at
higher redshifts, far from dark energy domination at low $z$  (see
Fig.~\ref{fig_iswrs_z}). 
This excess somehow balances the suppression in the
weak-lensing signal, and was first pointed out by
\cite{Lesgourgues_etal_2008} for the case of the ISW-galaxy
cross-correlation. Similar arguments explained in their \S~{\bf B}
hold also in our case. It has been also shown \cite{Schaefer2011} that
ideal cosmic variance limited experiments may detect, via
cross-correlation with lensing, the non-linear RS effect with a
significance of $\sim 3 \sigma$ integrating up to multipoles
$\ell=3\times10^3$. However, this significance is drastically reduced by the finite resolution
and noise of actual CMB experiments, so that the signal to noise ratio of the
non-linear RS effect results to be an order of magnitude smaller compared to that of
the linear ISW effect. This implies that it would be quite difficult to
measure it with present and near future CMB-LSS experiments. 

Together with the results presented in \S~\ref{ISWRS--CMB-lensing
  cross}, these represent the main findings of the present work. For the
first time in the literature, the cross-correlation between CMB/weak-lensing
and ISW-RS effects have been simulated, on a very large range of scales,
from the linear to the fully non-linear regimes, and in the presence of
massive neutrinos.

\section{Conclusions}
\label{conclu}
In this work we present full-sky maps, auto and cross angular spectra of the ISW-Rees-Sciama and
CMB/weak-lensing signals, from the linear to the fully non-linear
regimes, as extracted via direct ray-tracing across very large N-body
simulations including a massive neutrino component, the so-called
DEMNUni simulations. We assume a Planck-like baseline cosmology, and
add neutrinos with total masses $M_\nu=0,\,0.17,\,0.3,\,0.53$ eV,
fixing the normalisation of the matter power spectrum at CMB.

The analysis of these signals shows that

\begin{itemize}
\item
Free-streaming massive neutrinos induce a time variation, $\dot{\Phi}$,
of the gravitational potential, affecting mostly scales corresponding to the
transition from the linear to the non-linear regimes, $l\sim 100$, and producing
a non negligible contribution to the ISW effect (see Fig.~\ref{fig_iswrs}). The induced
$\Delta$T anisotropies are more important for light neutrino masses and
at high redshifts $z\gtrsim 1.5$, when dark energy is not the dominant component. At lower
redshifts the ISW effect becomes more suppressed as the neutrino mass
increases (Fig.~\ref{fig_iswrs_z}). Considering relative differences with respect to the massless
case, on linear scales we recover,
within $\sim 1-2$\%, accuracy, analytical expectations from \CAMB,
which, however, at the moment does not provide non-linear estimations of the ISW-RS signal. 

\item
At non-linear scales, $l>200$, massive neutrinos decrease the angular
power spectrum corresponding to the Rees-Sciama effect. Such
suppression is larger for larger neutrino masses, and decreases with
increasing redshifts, as shown in Fig.~\ref{fig_iswrs_z}.

\item
Since lensing traces directly the matter power spectrum, which can be
largely suppressed for large neutrino masses, we recover a similar
suppression in the CMB- and weak-lensing signals (Figs.~\ref{fig_cmblens}-\ref{fig_weaklens}).
When looking at relative differences with respect to the massless
case, the agreement with \CAMB\ (non-linear neutrino
corrections included) is at $\sim 1\%$ level on the scales covered by the simulations.
The suppression of the lensing auto power spectra decreases with the
increase of the source redshift, $z_s$, with a maximum difference of
$\sim 10\%$ between $z_s=2$ and $z_s=1100$.

\item
Lensed TT, EE and BB spectra are consistently affected by massive
neutrinos (mid and lower panels of Fig.~\ref{fig_cmblens}). As
expected, since the strength of the
gravitational potential decreases for larger neutrino masses, CMB
acoustic oscillations are less smoothed and smeared out than in the
massless case. This implies that, as the neutrino mass increases, the lensed TT and EE power spectra are larger at small $l$, and the
so-called ``damping-tail'' is lower at high $l$.
Therefore the amplitude of the lens-induced B-mode power
spectrum decreases for larger neutrino masses  on all the
scales, as shown in the left lower panel of Fig.~\ref{fig_cmblens}.

\item
Concerning the cross-correlation between ISW-RS and CMB-lensing
signals, at $l\lesssim 400$ we correctly recover the linear signal
from \CAMB, within $1-2 \%$ accuracy
(Fig.~\ref{fig_cmbl_isw_cross}). At the transition between the linear
and the non-linear regimes, $l\sim 700$, the simulated signal
correctly undergoes the sign-inversion expected by non-linear
semi-analytical calculations in the massless case \cite{Schaefer2011,Mangilli13, Lewis_RS}. This feature is
interestingly altered by the presence of massive neutrinos, as it moves
toward larger multipoles with increasing $M_\nu$, and the
displacement, with respect to the $M_\nu=0$ eV case, can be larger than a factor of
$2$ for $M_\nu=0.53$ eV.  This is reasonably
expected, since massive neutrinos extend the linear regime to smaller
scales than in the massless case. Therefore, the cross power between
ISW-RS and CMB-lensing increases with larger neutrino masses at
$300\lesssim l \lesssim 1500$, and, {\emph e.g.}, at $l\sim 600$, we find an excess of cross
power of a factor of $\sim 4$ for $M_\nu=0.3$ eV.   
At higher multipoles, in the fully non-linear regime, we find a
suppression of the signal as also occurs for the auto spectra.

\item
As shown in the left panel of Fig.~\ref{fig_weakl_isw_cross}, the cross-correlation
between ISW-RS and weak-lensing presents features similar to
the ISW-RS cross CMB-lensing signal, with the only difference given by a
lower total impact of massive neutrinos. Again, when looking at relative differences with respect to the massless
case, at $l\lesssim 300$ we correctly recover,  within $1-2 \%$ accuracy, the linear
signal from \CAMB, which does not provide non-linear estimations at the moment. 
The non-linear sign-inversion of the cross power is still present but less enhanced, while the
signal seems to increase with increasing $z_s$. In particular, the
excess of ISW-RS due to the presence of massive neutrinos makes the
cross power less suppressed with respect to the weak-lensing auto
power, and finally, for very high source redshifts, we observe a net
excess of power with respect to the massless case (right panel of Fig.~\ref{fig_weakl_isw_cross}).
\end{itemize}  

The last two points represent the main findings of this work. 
The cross-correlation between the ISW-RS and lensing signals enters the
computation of the lensed CMB temperature bispectrum~\cite{Lewis_RS}; therefore its
correct  estimation at the non-linear level from N-body simulations,
and the knowledge of the neutrino impact on its amplitude may result to be
of extreme importance for the full evaluation of the CMB
temperature three-point function~\cite{Lewis2011}. 
The latter probes the perturbation growth and expansion history of the
Universe, and hence can be 
used to constrain dark energy and neutrino masses~\cite{Lewis06}. In addition, due to
non-linear structure evolution, on very small scales the lensing
potential is not a Gaussian field,  and consequent additional
contributions 
to the bispectrum may be evaluated directly from the simulated maps
obtained via the DEMNUni simulations. We reserve this for future work.

\acknowledgments
C.C. thanks Anna Mangilli, Julien Bel, Emiliano Sefusatti, Matteo Calabrese,
and  Matteo Zennaro for very useful discussions.
C.C. thanks Matteo Viel for providing the N-GenIC code for
initial conditions, modified to take into account a massive neutrino
particle component. The DEMNUni simulations were carried out 
at the Tier-0 IBM BG/Q machine, Fermi, of the Centro Interuniversitario del
Nord-Est per il Calcolo Elettronico (CINECA, Bologna, Italy), via the five
million cpu-hrs budget provided by the Italian SuperComputing Resource
Allocation (ISCRA) to the class--A proposal entitled ``The Dark Energy and
Massive-Neutrino Universe".  C.C. acknowledges financial support from the INAF
Fellowships Programme 2010 and from the European Research Council through
the Darklight Advanced Research Grant (n. 291521).
K.D. and M.P. acknowledge support by the DFG Cluster of
Excellence ``Origin and Structure of the Universe'' and the
SFB-Tansregio TR33 ``The Dark Universe''.

\bibliographystyle{JHEPb}
\bibliography{CPD2016_jcap}

\end{document}